\documentclass[fleqn,usenatbib,onecolumn]{mnras}
\usepackage{natbib}
\usepackage{apjfonts}
\usepackage{amsmath}
\usepackage[table]{xcolor}

\usepackage{epsfig}
\usepackage{graphicx}
\usepackage{amssymb}
\usepackage{aas_macros}
\usepackage{color}

\newcommand{\kms}{\,{\rm km \, s^{-1}}}
\newcommand{\AU}{\,{\rm AU}}
\newcommand{\pc}{\,{\rm pc}}
\newcommand{\kpc}{\,{\rm kpc}}
\newcommand{\msol}{\,M_\odot}

\newcommand{\bb}{\mathbfit }
\newcommand{\oversim}[2]{\protect{\mbox{\lower0.5ex\vbox{%
   \baselineskip=0pt\lineskip=0.2ex
   \ialign{$\mathsurround=0pt #1\hfil##\hfil$\crcr#2\crcr\sim\crcr}}}}} 
\catcode`\"=\active\let"=\"  

\def\3{{\ss} }

\def\c12{{1\over 2}}   
\def\erf{{\rm erf}}   
   
\def\d{{\rm d}}   
   
\def\plusplus{\raise 0.3ex\hbox{${\scriptstyle ++}$}{}}

\def\and{{{\rm M}31}}

\def\myr{\,{\rm Myr}}

\bibliography{biblio}

\usepackage{array}
\newcolumntype{L}[1]{>{\raggedright\let\newline\\\arraybackslash\hspace{0pt}}m{#1}}
\newcolumntype{C}[1]{>{\centering\let\newline\\\arraybackslash\hspace{0pt}}m{#1}}
\newcolumntype{R}[1]{>{\raggedleft\let\newline\\\arraybackslash\hspace{0pt}}m{#1}}

\begin{document}
\title[Trapped matter in the solar system]{A halo of trapped interstellar matter surrounding the solar system}
\author[Jorge Pe\~{n}arrubia]{Jorge Pe\~{n}arrubia$^{1,2}$\thanks{jorpega@roe.ac.uk}\\
  $^1$Institute for Astronomy, University of Edinburgh, Royal Observatory, Blackford Hill, Edinburgh EH9 3HJ, UK\\
  $^2$Centre for Statistics, University of Edinburgh, School of Mathematics, Edinburgh EH9 3FD, UK
}
\maketitle  

\begin{abstract}
This paper shows that gravitating bodies travelling through the Galaxy can trap lighter interstellar particles that pass nearby with small relative velocities onto temporarily-bound orbits.
  The capture mechanism is driven by the Galactic tidal field, which can decelerate infalling objects to a degree where their binding energy becomes negative.
  Over time, trapped particles build a local overdensity -- or `halo'-- that reaches a steady state as the number of particles being captured equals that being tidally stripped.
   This paper uses classical stochastic techniques to calculate the capture rate and the phase-space distribution of particles trapped by a point-mass.
   In a steady state, bound particles generate a density enhancement that scales as $\delta(r)\sim r^{-3/2}$ (a.k.a `density spike') and follow a velocity dispersion profile $\sigma_h(r)\sim r^{-1/2}$.
  Collisionless $N$-body experiments show excellent agreement with these theoretical predictions within a distance range $r\gtrsim r_\epsilon$, where $r_\epsilon\simeq 0.8\,\exp[-V_\star^2/(2\sigma^2)]\,Gm_\star/\sigma^2$ is the thermal critical radius of a point-mass $m_\star$ moving with a speed $V_\star$ through a sea of particles with a velocity dispersion $\sigma$.
  Preliminary estimates that ignore collisions with planets and Galactic substructures suggest that the solar system may be surrounded by a halo that contains the order of $N^{\rm ISO}(<0.1\pc)\sim 10^7$ energetically-bound 'Oumuamua-like objects, and a dark matter mass of $M^{\rm DM}(<0.1\pc)\sim 10^{-13}M_\odot$.
The presence of trapped interstellar matter in the solar system can affect current estimates on the size of the Oort Cloud, and leave a distinct signal in direct dark matter detection experiments.
\end{abstract}

\section{Introduction}\label{sec:intro}
The solar system consists of the Sun and the population of planets and minor bodies gravitationally bound to it. Although this definition is unambiguous, it ignores the Milky Way. To date, the dynamical interplay between the Sun and the interstellar surroundings is poorly understood, specially in the outer-most regions of the solar system, where external forces become comparable to the Sun's gravitational attraction. In this region, the boundary with the Galaxy is blurred, as there is a continuous exchange of material that goes in both ways: from the interstellar medium into the solar system via gravitational capture, and from the solar system out to the Milky Way via tidal heating and stripping\footnote{In this paper, the word `capture' is broadly used to describe any dynamical process wherein a body undergoes a transition from Milky Way orbit to a heliocentric one.}.

Most planet formation models assume that the solar system formed about 4.5 billion years ago in a dense molecular cloud. Planets, moons, asteroids, and other small bodies formed in a protoplanetary disc that became progressively eroded due to collisions with the planets and the action of external tidal forces (e.g. Kokubo \& Ida 1998; Portegies Zwart 2021a and references therein). In this scenario, the material currently orbiting the Sun represents the surviving remnants of a nearly obliterated disc (e.g. Charnoz \& Morbidelli 2003; Johansen et al. 2021, Portegies Zwart 2021; Portegies Zwart et al. 2021).

This picture has been recently challenged by the discovery of two visitors from the interstellar space, the 'Oumuamua object (Meech et al. 2017) and the comet Borisov (Jewitt \& Luu 2019), which provided undeniable evidence that the solar system may also host a sizeable population of minor bodies of extra-solar origin (e.g., Siraj \& Loeb 2019; Namouni \& Morais 2020). 
Like 'Oumuamua and Borisov, the vast majority of interstellar objects are expected to move across the solar system on unbound trajectories.
Although these visitors from outer space represent transient occurrences, they point to the possibility that the Sun may have captured an unknown number of these objects in the past. 

It is important to bear in mind that gravitational capture cannot happen in the classical two-body problem. Consider for example an interstellar body approaching the Sun on a hyperbolic, unbound trajectory with a specific energy $E_\infty=v^2_\infty/2> 0$, where $v_\infty$ is the asymptotic relative velocity in the limit $t\to -\infty$. If there is no internal or external force acting on this object other than the gravitational attraction of the Sun, then energy conservation demands $E=v^2/2-Gm_\star/r=E_\infty$. This in turn means that the infalling speed progressively accelerates as it approaches the Sun as $v^2=v^2_\infty+2Gm_\star/r$, which is typically known as `gravitational focusing', and experiences a mirrored deceleration on its way out of the solar system. 
Therefore, gravitational capture is not possible unless one or several mechanisms `slow down' the infalling body and dissipate its kinetic energy to a degree wherein the binding energy becomes negative, $E<0$.

The best studied dissipative mechanism for gravitational capture is 3-body interactions with planets (e.g., Valtonen \& Innanen 1982; Valtonen 1983; Siraj \& Loeb 2019, 2021; Lingam \& Loeb 2018; Hands \& Dehnen 2020; Napier et al. 2021a,b; Dehnen \& Hands 2022; Dehnen et al. 2021).
Dark Matter (DM) particles --if they exist-- can also be captured out of the Milky Way DM halo via collisions with planets. This scenario has been extensively explored for the case of the solar system (Gould 1987, 1988; Lundberg \& Edsj\"o 2004; Xu \& Siegel 2008; Peter 2009a,b,c; Edsj\"o \& Peter 2010; Iorio 2010; Anderson et al. 2020; Lehmann et al. 2021). More recently, Moro-Mart\'in \& Norman (2022) have also studied the effect of gas drag on the capture of interstellar objects during the early stages of star and planet formation.

In past studies, capture rates by the solar system are typically calculated under the assumption of {\it isolation} (i.e. neglecting external forces), and sampling a distribution of encounter speeds that follows either the velocity distribution of Milky Way stars in the Solar neighbourhood, or that appropriate for the solar birth environment (e.g. Portegies Zwart 2009; Adams 2010; Pfalzner 2013; Parker 2020).

To date, the role of the Milky Way potential in capturing interstellar matter remains poorly understood. To address this problem, it becomes necessary to drop the isolation assumption and account for the motion the Sun and that of the interstellar visitors around the Milky Way centre.
To this aim, two sets of coupled differential equations must be solved simultaneously. The first one describes the orbit of the Sun around the Milky Way
 \begin{align}\label{eq:eqmotsun}
 \ddot {\bb R_\star}=-\nabla\Phi_g({\bb R_\star}).
 \end{align}
 where ${\bb R_\star}$ is the galactocentric location of the Sun and $\Phi_g$ is the Galactic potential. The second set describes the trajectories of interstellar objects in the vicinity of the solar system 
 \begin{align}\label{eq:eqmot}
   \ddot {\bb R}=-\nabla\Phi_g({\bb R})+{\bb F}_\star+{\bb F}_p.
 \end{align}
 where ${\bb F}_\star=-Gm_\star\,({\bb R}-{\bb R}_\star)//|{\bb R}-{\bb R}_\star|^3$ is the specific force induced by the Sun, and ${\bb F}_p=-\sum_{i=1}^{N_p} G m_{p,i} ({\bb R}-{\bb R}_i)/|{\bb R}-{\bb R}_i|^3$ is the total specific force generated by $N_p$ planets (or any other massive bodies) moving in or around the solar system. For simplicity, the remainder of this work purposely sets $F_p=0$. We come back to this issue in Section~\ref{sec:discussion}. Note also that the `collisionless' approximation corresponds to the case $F_\star=F_p=0$, wherein the orbits of interstellar objects are solely governed by the smooth galactic potential $\Phi_g$.

 \begin{figure}
\begin{center}
\includegraphics[width=164mm]{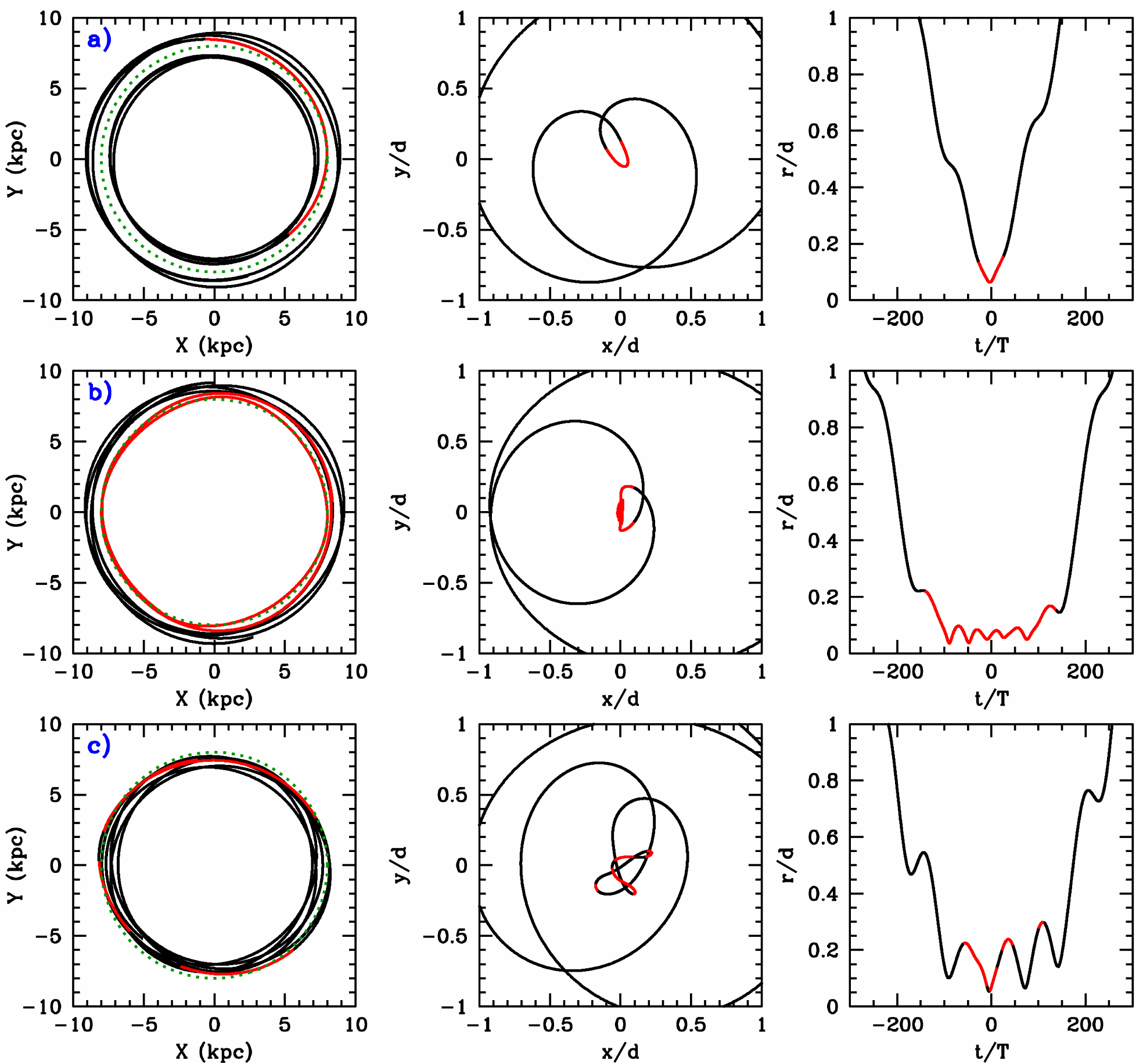}
\end{center}
\caption{Trajectories of particles trapped around a point mass $m_\star/M_g=5.3\times 10^{-4}$ moving on a circular orbit with galactocentric radius $R=8\kpc$ (dotted-green line), where $M_g=1.84\times 10^{12}M_\odot$ is the mass of the host dark matter halo (go to \S\ref{sec:tests} for details on the numerical experiments). Left and middle columns show orbits in a reference frame centred at the host Galactic and Keplerian potentials, respectively.
  Black/red solid lines show the orbits of trapped particles with positive/negative orbital energies $E=v^2/2-Gm_\star/r$. The separation between the point-mass and the dark matter particle is plotted in the right column as a function of time in units of the scale-length $d\equiv |\nabla \rho/\rho|^{-1}$. 
Trajectories can be broadly divided in three categories: a) Transient captures have a duration comparable to the local dynamical time, $T(d)$, given by~(\ref{eq:Ta}). (b) Semi-stable captures perform several revolutions in the Keplerian potential before being tidally stripped. (c) Repeated captures move on intricate trajectories dictated by the combined Sun+Galactic potentials.}
\label{fig:xyz}
 \end{figure}
 
 Defining the relative separation vector ${\bb r}={\bb R}-{\bb R}_\star$, and Taylor expanding $\nabla\Phi_g$ around the
 solar location at leading order $\mathcal{O}(r/R)$ yields a single differential equation for the relative
separation between the pair, which is known as the {\it tidal approximation}
\begin{align}\label{eq:eqmottide}
\ddot{\bb r}\equiv \ddot{\bb R}-\ddot{\bb R}_\star\approx -\frac{G m_\star}{r^3}{\bb r}+{\bf T}\cdot {\bb r};
\end{align}
where
 \begin{align}\label{eq:tensor}
T^{ij}=-\frac{\partial^2\Phi_g}{\partial x_i\partial x_j}\bigg|_{{\bb R}_\star},
\end{align}
 is a 3x3 {\it tidal tensor}. The presence of a tidal field induces a velocity variation
 \begin{align}\label{eq:delta_v}
   \Delta {\bb v}=\int_S \d t\,{\bb F}_{\rm tide}(t),
   \end{align}
 where ${\bb F}_{\rm tide}={\bf T}\cdot {\bb r}$ is the tidal force, and $S$ denotes the trajectory of an infalling object from $t\to -\infty$ to the present. Hence, in general the orbital energy of infalling Milky Way particles is not a conserved quantity 
\begin{align}\label{eq:delta_E}
 \Delta E=E-E_\infty=\Delta {\bb v}\cdot{\bb v}+\frac{1}{2}|\Delta {\bb v}|^2.
\end{align}
here, $E=v^2/2-\Phi_\star$ is the binding energy in a Keplerian potential $\Phi_\star=-Gm_\star/r$, and $r$ and $v$ are the relative distance and velocity to the point-mass $m_\star$.
Crucially, Equations~(\ref{eq:delta_v}) and~(\ref{eq:delta_E}) show that, depending on their individual trajectories, Galactic tidal forces can either accelerate ($\Delta E>0$) or decelerate ($\Delta E<0$) infalling bodies approaching the point-mass $m_\star$. This mechanism is therefore akin to slingshot manoeuvres, which can be designed to increase/decrease the speed of a spacecraft or redirect its path. 
In this paper, we are interested in objects that lose enough kinetic energy as to become bound to the solar system, such that $E<0$. For clarity, we shall refer to this particular type of capture process as {\it tidal trapping}. 
The questions we aim to answer are the following: how many interstellar bodies are expected to be tidally trapped on bound orbits around the Sun at present? what is their baryon-to-dark matter ratio? how are they distributed across the Solar system? how long do they remain energetically bound?

Unfortunately, tidally-trapped particles move on very intricate trajectories in the proximity of the Sun, hindering an analytical approach to these questions. For illustration, Fig.~\ref{fig:xyz} shows the motion of three particles that become temporarily bound to a massive point-mass moving on a circular orbit around the host galaxy potential (see \S\ref{sec:tests} for details on the numerical calculations). All trapped particles found in our numerical analysis remain bound to the Keplerian potential for a limited amount of time and therefore represent transient captures.
We also found that the trajectories of these objects can be broadly separated in three categories:
a) {\it Transient capture}. In the majority of capture events, interstellar particles only remain bound to the point-mass $m_\star$ for the duration of a single encounter. This is illustrated in the top panels of Fig.~\ref{fig:xyz}, which colours points along the trajectory with negative and positive binding energies with red and black lines, respectively. Top-left panel shows a particle with an original galactocentric apocentre of $\sim  9\kpc$ being decelerated during a single interaction with $m_\star$. As result, orbital energy is lost and the particle is transferred to a new galactocentric orbit with apocentre $\sim 7\kpc$. The complexity of the fly-by approach is highlighted in the middle panel, which shows the trajectory in the reference frame of the point-mass $m_\star$. Notice that the binding energy only turns negative close to perihelion ($t=0$), suggesting that the same dynamical mechanism that allows this particle to lose kinetic energy in the first place is also responsible for its later unbinding.
b) {\it Semi-stable capture} (middle row). In this case, tidally-trapped particles undergo several revolutions around the point-mass $m_\star$ before being released back to the Galactic potential. These objects are particularly interesting given that the probability of collisions with planets is proportional to the time that an interstellar particle spends in the solar system. 
c) {\it Repeated capture} (bottom row). The most convoluted trajectories corresponds to particles that move in the Galactic potential and experience periodic interactions with the point-mass $m_\star$. The sign of the binding energy can be seen to alternate between positive and negative values within short time intervals.
A crucial result highlighted in Fig.~\ref{fig:xyz} is that tidal trapping {\it always} lead to temporary captures. We study to this important aspect below with the aid of numerical experiments.

 \begin{figure*}
\begin{center}
\includegraphics[width=164mm]{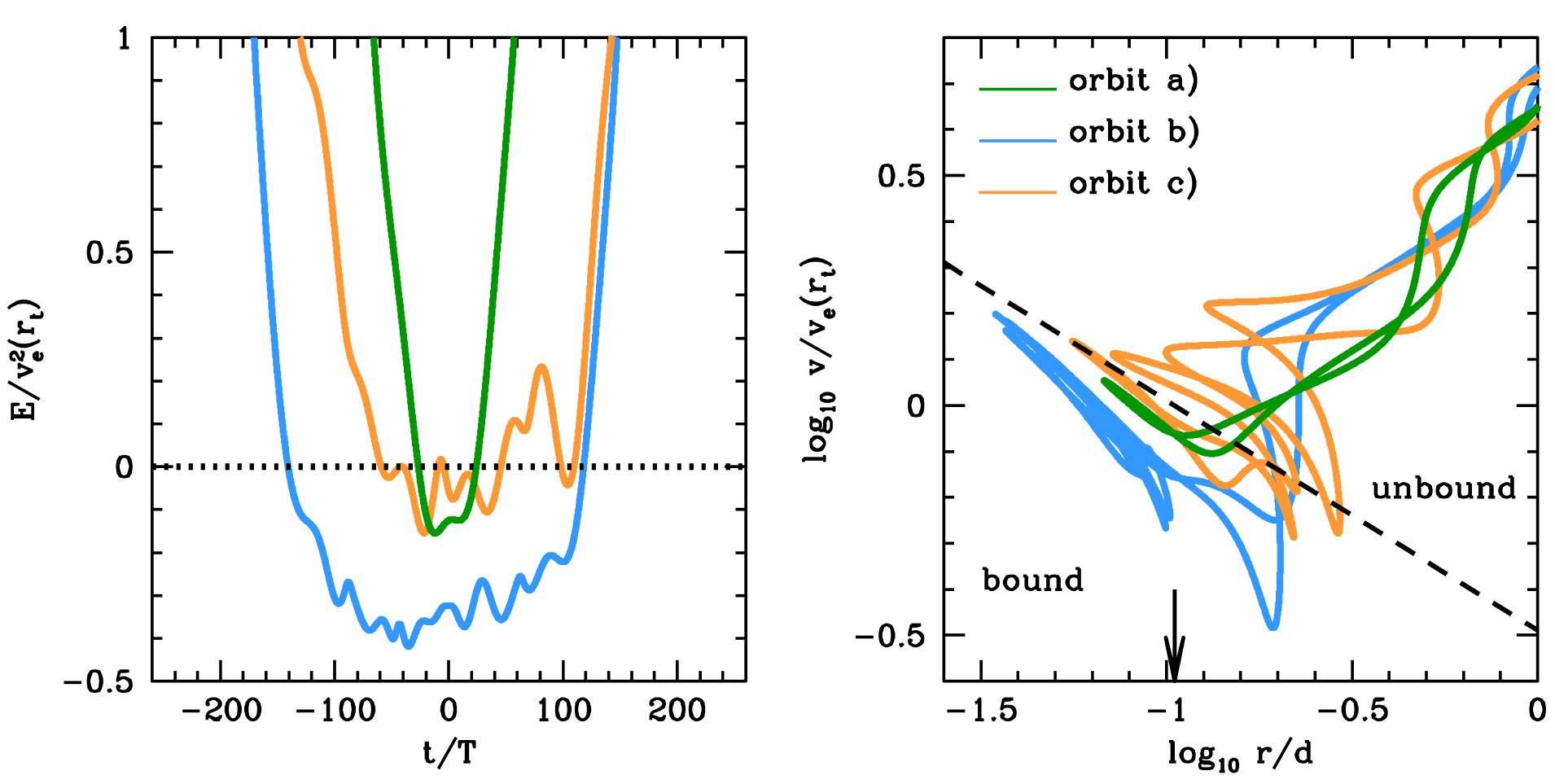}
\end{center}
\caption{{\it Left panel}: Specific binding energy, $E=v^2/2-Gm_\star/r$, as a function of time for the particles shown in Fig.~\ref{fig:xyz}. Energy is measured in units of the escape velocity $v_e(r)=\sqrt{2Gm_\star/r}$ measured at the tidal radius $r=r_t$, Equation~(\ref{eq:rt_g}), whereas time is normalized by the fluctuation mean-life, $T(d)$, Equation~(\ref{eq:Ta}) (see text). {\it Right panel}: Relative distance and velocity of the three orbits plotted in the left panel. The dashed line marks the value of the escape velocity as a function of distance, and the vertical arrow shows the location of the tidal radius. Points above/below the dashed line have positive/negative binding energies. In this diagram, particles approach the point-mass from the upper-right corner, and leave the system following a similar trajectory. Note that particles with a constant energy move parallel to the dashed line (a.k.a `gravitational focusing' effect).  }
\label{fig:e}
 \end{figure*}

For a better understanding of the mechanics of tidal injection it is useful to examine how orbital energy varies during the capture process. This is plotted in the left panel of Fig.~\ref{fig:e} for the three trajectories shown in Fig.~\ref{fig:xyz}. The first noteworthy result is the confirmation that binding energy $E$ can {\it not} be assumed to remain constant during capture. Indeed, all trapped particles show very strong variations of energy as they approach the point-mass ($E$ sign becomes negative), and after they are released back to the Galactic potential ($E$ sign becomes positive).
The variation of energy is much smaller after the injection onto the Keplerian system has taken place, but even then it does not remain constant.
 As expected, the amount of time that a particle remains bound to the point mass $m_\star$ scales directly with the binding energy attained during injection. In particular, the trajectories of weakly-bound particles a) (`transient capture') and c) ('repeated capture') remain bound for a shorter time span than b) ('semi-stable capture'), which reaches much higher binding energies.

 Where does capture take place? To answer this question, we plot in the right panel of Fig.~\ref{fig:e} the phase-space coordinates of trajectories shown in Fig.~\ref{fig:xyz}, where $r$ and $v$ respectively denote distances and velocities relative to the point mass $m_\star$. In this diagram, interstellar particles approach $m_\star$ from the upper-right corner, and leave the Keplerian system approximately following the inverse trajectory.
 At small separations, trapped particles display an anticorrelation between distance and velocity expected from gravitational focusing. However, at large separations the relative distance {\it and} the relative velocity decrease as particles move towards the point-mass, which is not what one would naively expect from gravitational focusing. Indeed, if $E$ were a conserved quantity the relative velocity should converge towards an asymptotic limit $v_\infty$ at $r\to \infty$ as $v^2=v^2_\infty+2 Gm_\star/r$.
 In contrast, Fig.~\ref{fig:e} shows shows no asymptotic behaviour, which casts doubts on the physical meaning of $v_\infty$ for systems that are not in isolation. 
 The transition between these two regimes roughly happens at the tidal radius of the point-mass, $r_t$ (marked with a vertical arrow, see \S\ref{sec:tests} for details on its derivation).

It is important to stress that interstellar particles cannot visit the area under the escape velocity curve, $v<v_e$ (black-dashed line) if the solar system is assumed to be in isolation. The key difference in our models is therefore the galactic potential, which sources a tidal field that actively injects \& removes interstellar particles from this region of phase-space\footnote{From the point of view of statistical mechanics, tidal forces act as a `cooling' mechanism that lowers the kinetic energy of infalling interstellar objects. Notice, however, that heat transport goes in the opposite direction to what is expected in fluids and gases, i.e. from the relatively cold Keplerian system towards a relatively hot interstellar environment. This is due to the negative specific heat of gravitating systems (Antonov 1961; see Lynden-Bell 1999 for a review).}, as illustrated in Fig.~\ref{fig:e}.

Equations~(\ref{eq:eqmotsun}) and~(\ref{eq:eqmot}) correspond to a special case of the restricted three-body problem, where the most massive body (the host Galaxy) has an extended mass distribution. Capture processes in the `classical' three-body problem, where the three bodies are represented as point-masses, have been studied in the past. For example, the pioneer work of Szebehely (1967) showed that a finite number of solutions exists where the lightest object is transferred from one distinct mode of motion around the most massive body to another distinct mode around the intermediate-mass one. This mechanism was used by Hunter (1967), Heppenheimer (1975) and Heppenheimer \& Porco (1976) to study the formation of Jupiter's outer satellites, and by Singer (1968) to analyze the origin of the Moon as a planetoid captured by the Earth. More recently, Suetsugu \& Ohtsuki (2013) investigate the general case of temporary capture of planetesimals by a giant planet, while J{\'\i}lkov{\'a} et al. (2015) explore the scenario where the inner Oort Cloud was captured from another star during a close encounter in their birth cluster. 
Following up on the above results, Higuchi \& Ida (2017) analyze capture of asteroids by a planet moving on an eccentric orbit, finding that temporary capture becomes more difficult as the planet's eccentricity increases. Daniel et al. (2017) analyzed the opposite process, namely the escape of energetically-unbound particles from their progenitor stellar cluster. It was found that a number of ``potential escapers'' remained within two Jacobi radii for several dynamical periods, even though their binding energy remained positive during this time. Recently, three-body captures in accretion discs have also gained attention as a possible source of Black-Hole binaries. E.g Li et a. (2022) and Boekholt et al. (2022) show that close encounters between Black Hole pairs moving on circular disc orbits around a supermassive Black Hole can form bound pairs with the help of gravitational wave dissipation. 

Here, it is worth highlighting the work of Petit \& Henon (1986), who used numerical method to follow the interaction of two small satellites, both initially moving on circular orbits around a Keplerian potential. Two important results in that paper are relevant in our analysis: first, their work confirmed earlier claims that captures in a three-body system are temporary events that ultimately end up the dissolution of the bound pair. Furthermore, it was found that capture only happens for very precise combinations of impact parameters and relative velocities which exhibit a self-similar, Cantor-like structure. More recent work of Boekholt et al. (2022) confirmed these results and found that the phase space structure that leads to capture resembles a Cantor set with a fractal dimension of 0.4.


Taking these results at face value portrays the outer solar system as a dynamically active region full of interstellar objects trapped on intricate, unstable orbits that cross the tidal radius repeatedly and blur the boundary between the solar system and the Milky Way.
Alas, following the capture and subsequent escape of tidally-trapped particles via numerical solutions to the equations of motion~(\ref{eq:eqmotsun}) and~(\ref{eq:eqmot}) quickly becomes a CPU-demanding task when one considers the astronomically-large number of minor bodies populating the interstellar space.

As an alternative, this paper shows that a deeper theoretical understanding of ensemble properties of tidally-trapped bodies can be obtained by abandoning the classical approach of solving deterministic equations of motion for individual objects, and focusing instead on a statistical description of the phase-space distribution of interstellar particles in the vicinity of a point-mass $m_\star$. 
Section~\ref{sec:stats} introduces a statistical theory that allow us to compute the number of tracer particles trapped by a point-mass $m_\star$ as it travels through an extended galaxy, and derive their steady-state distribution in the Keplerian potential $\Phi_\star(r)=-Gm_\star/r$.
Section~\ref{sec:tests} shows a number of numerical experiments devised to test the theory. In these experiments, a massive point-mass $m_\star$ captures tracer particles from a spherical, non-rotating halo initially in dynamical equilibrium.
Section~\ref{sec:discussion} discusses the implication of our findings in the context of the solar system. In particular, we estimate the number of Interstellar Objects (ISOs) and DM particles trapped on bound orbits around the Sun.

\section{Statistical theory}\label{sec:stats}

This Section presents a statistical theory that allows us to (i) estimate the  umber of particles that become gravitationally bound to a point-mass $m_\star$ (the ``Sun'') as it travels through a galaxy made of lighter particles with individual masses $m\ll m_\star$ (the ``galaxy''), as well as (ii) derive their steady-state distribution in the Keplerian potential $\Phi_\star(r)=-Gm_\star/r$.

Analytical solutions to this problem can be found using classical stochastic techniques insofar as field particles in the vicinity of $m_\star$ can be assumed to follow random trajectories. As shown below, this condition is met in the {\it collisionless limit}, in which field particles move on unperturbed orbits around a smooth Galactic potential $\Phi_g$.
From a statistical framework, the collisionless approximation is accurate insofar as the number of particles affected by $m_\star$ represents a very small fraction of the total number of field particles located within a volume element $V$ centred around the point-mass at any given time.
We will see that this condition requires the local velocity dispersion of the host galaxy ($\sigma$) to be much higher than the escape speed of the Keplerian potential ($v_e$), i.e. $v_e\ll \sigma$, such that adding or removing the point-mass $m_\star$ leads to negligible effects on the local distribution function of the host galaxy.

For clarity, the problem is broken in three parts. First, in \S\ref{sec:bound} derive an analytical equation for the number of field particles with relative velocities $v<v_e$ within a volume $V$ around point-mass ($N_b$).

Second, in \S\ref{sec:acc} we assume that particles move on statistically uncorrelated (random) trajectories, which demands $N_b\ll N=n\,V$, where $n$ is the local number density of field particles. This condition allows us to use Smoluchowski (1916) statistical theory to derive an analytical expression for the rate at which field particles enter $V$ with negative binding energies, $E<0$.

The third and most difficult aspect of the analysis is to account for the impact of $m_\star$ on the trajectories of infalling objects. To this aim, \S\ref{sec:survival} develops a statistical description of the limited life time of tidally-trapped particles in the Keplerian potential $\Phi_\star$, i.e. the time over which a trapped particle continuously has $E<0$ before being lost to galactic tides. This in turn can be used to estimate the number of bound interstellar particles in steady-state ($N_{\rm ss}$). We will see that that $N_{\rm ss}=\alpha\,N_b$, where $\alpha$ is a dimension-less, order-unity parameter that is empirically computed from the distribution of survival times in Section~\ref{sec:tests}. Lastly, \S\ref{sec:profile} derives the equilibrium distribution of field particles with $E<0$ orbiting around $\Phi_\star$.

We will also study the limitations of the theory in detail. Of particular relevance is the singular behaviour of the Keplerian forces at $r=0$, which translates in to a escape velocity that diverges $v_e=2Gm_\star/r\to \infty$ in the limit $r\to 0$. This shortcoming is analyzed in Section~\ref{sec:tests} with the aid of $N$-body experiments.

\subsection{Bound particles in the vicinity of the point-mass}\label{sec:bound}
Let us first estimate the number of background particles expected to be found in a volume element centred at the point-mass ($V$) with relative velocities below the escape speed, $v<v_e$
This is done under two assumptions: (i) field particles move on uncorrelated trajectories, and (ii) their distribution in phase-space follows the local distribution function of the host galaxy.
Both conditions are valid in the {\it collisionless limit}, where the presence of $m_\star$ induces negligible effects on the local galactic background\footnote{Under this condition, the boundary $E=0$ loses any dynamical meaning and the problem becomes purely statistical, i.e. it reduces to counting particles with relative distances ($r$) and velocities ($v$) such that $v<v_e(r)$ (or $E < 0$), which we use as the main criterium for capture.}.
We will re-visit this approximation in Section~\ref{sec:survival}, which accounts for the effect of the gravitational attraction of $m_\star$ on the motion of bound particles.

For simplicity, in what follows the host galaxy is assumed to be made of light particles ($m\ll m_\star$) in a state of dynamical equilibrium, such that the number density can be derived from the constant local matter density as $n(\bb R)=\rho(\bb R)/m$. In our notation, phase-space quantities measured in the galaxy frame are shown with capital letters, and those measured with respect to the point-mass with lower letters.

The size of the volume element centred at $\bb R_\star$ is to be sufficiently small to guarantee that $r\ll d(\bb R_\star)\equiv |\nabla \rho/\rho|^{-1}_{\bb R_\star}$, such that the number density can be assumed to be approximately constant, i.e. $n({\bb R_\star}+{\bb r})\approx n({\bb R_\star})\equiv n$ for $r\ll d$. Hence, the distribution of background particles becomes homogeneous across $V$, i.e. $p(\bb r)\d^3 r\approx 4\pi r^2 \,n\,\d r$ at $r\ll d$, which is known as the {\it local approximation}.
 
The relative velocity distribution of particles within the volume element $V$ is assumed to follow a Maxwellian distribution displaced by the reflex velocity of the point-mass 
\begin{align}\label{eq:pv}
  p(\bb v)=\frac{1}{(2\pi \sigma^2)^{3/2}}\exp\bigg[-\frac{(\bb v+\bb V_\star)^2}{2\sigma^2}\bigg],
\end{align}
 where $\sigma=\sigma(\bb R_\star)$ is the local, one-dimensional velocity dispersion of the host galaxy. The mean squared (relative) velocity between the background particles and the point-mass $m_\star$ can be straightforwardly derived from~(\ref{eq:pv}) as 
\begin{align}\label{eq:v2a}
  \langle v^2\rangle &= \int \d^3 v\,p(\bb v)\,v^2= 3\sigma^2 + V_\star^2.
\end{align}

Following Pe\~narrubia (2021, hereafter P21), the number of background particles within the volume $V$ with specific energy $E=v^2/2-Gm_\star/r<0$ can be can be calculated under the local approximation and Maxwellian approximations as
\begin{align}\label{eq:Nb1}
  N_b(r)&=\int_V\d^3 r\, n(\bb r) \int_{E<0}\d^3v\,p(\bb v) \\\nonumber
  &=\frac{8\pi^2n }{(2\pi\sigma^2)^{3/2}} \int_0^r\d r'\, r'^2\int_0^{v_{e}(r')}\d v\,v^2 \int_{-1}^{+1}\d x\, e^{-\frac{v^2+V_\star^2 +v V_\star x}{2\sigma^2}}\\\nonumber
  &=2\sqrt{2\pi}\frac{n }{\sigma^3} \int_0^r\d r'\, r'^2\int_0^{v_{e}(r')}\d v\,v^2 \frac{\sigma^2}{v V_\star}e^{-(v+V_\star)^2/(2\sigma^2)}\bigg(e^{2 vV_\star/\sigma^2}-1\bigg)\\ \nonumber
  &= 2\sqrt{2\pi} \frac{n}{\sigma^3} \int_0^r\d r'\, r'^2 \bigg\{ -\frac{\sigma^4}{V_\star}e^{-(v_e+V_\star)^2/(2\sigma^2)}\bigg(e^{2v_e V_\star/\sigma^2}-1\bigg) +\sqrt{\frac{\pi}{2}}\sigma^3\bigg[\erf\bigg(\frac{V_\star+v_e}{\sqrt{2}\sigma}\bigg)-\erf\bigg(\frac{V_\star-v_e}{\sqrt{2}\sigma}\bigg)\bigg] \bigg\}
\end{align}
here $\erf(x)$ is the error function, and $v_e(r)=\sqrt{2 G m_\star/r}$ is the escape speed. Using the condition $v_e\ll \sigma$, the integrand term within brackets can be approximated as
\begin{align}\label{eq:pv_app}
-\frac{\sigma^4}{V_\star} e^{-(v_e+V_\star)^2/(2\sigma^2)}\bigg(e^{2v_e V_\star/\sigma^2}-1\bigg) +\sqrt{\frac{\pi}{2}}\sigma^3\bigg[\erf\bigg(\frac{V_\star+v_e}{\sqrt{2}\sigma}\bigg)-\erf\bigg(\frac{V_\star-v_e}{\sqrt{2}\sigma}\bigg)\bigg]
=\frac{2}{3}e^{-V_\star^2/(2\sigma^2)}v_e^3  +\mathcal{O}\bigg(\frac{v_e}{\sigma}\bigg)^5.
\end{align}
Thus, at leading order Equation~(\ref{eq:Nb1}) becomes 
\begin{align}\label{eq:Nb}
  N_b(r)&\approx \frac{4\sqrt{2\pi}}{3} e^{-V_\star^2/(2\sigma^2)} \frac{n}{\sigma^3} \int_0^r\d r'\, r'^2 v_e^3(r')\\ \nonumber
  &=\frac{32\sqrt{\pi}}{9}(G m_\star)^{3/2}\frac{n}{\sigma^3}e^{-V_\star^2/(2\sigma^2)}r^{3/2}.
\end{align} 
which recovers Equation~(4) of P21 for the case of a point-mass at rest ($V_\star=0$). The number of bound particles within the volume $V$ is therefore proportional to the {\it phase-space density} of light background particles at the location of the point-mass, $Q_g\equiv Q_g(\bb R_\star)= n/\sigma^3$. The value of $N_b$ drops exponentially for point-masses travelling across the background ($V_\star\ne 0$).

The above formula plays an important role in the remainder of this work, and is accurate insofar as the local approximation holds, $r\ll d$, and the point-mass is sufficiently light, such that the escape velocity $v_e=(2Gm_\star/r)^{1/2}\ll \sigma$. This latter condition introduces a critical radius ($r_0$) at small distances where the escape speed becomes comparable to the velocity dispersion of the background particles. Setting $v_e(r_0)=\sigma$ yields $r_0=2Gm_\star/\sigma^2$. Therefore, the statistical theory is valid within the range $r_0\lesssim r\ll d$, with the lower limit approaching $r_0\to 0$ as $m_\star\to 0$.

Recall that the theory is built under the assumption that bound particles represent a negligible fraction of the total number of field particles enclosed within the same volume, $N=n\,V$. Using~(\ref{eq:Nb}) it is straightforward to show that the condition $v_e\ll \sigma$ meets this demand
\begin{align}\label{eq:NbN}
  \frac{N_b}{N}\bigg|_r=\frac{2\sqrt{2}}{3\sqrt{\pi}} \bigg(\frac{v_e}{\sigma}\bigg)^3e^{-V_\star^2/(2\sigma^2)} \lll 1 ~~~~~~{\rm for}~~~~\frac{v_e}{\sigma}\ll 1.
\end{align}

\subsection{Entrapment rate}\label{sec:acc}
Next goal is to compute the number of particles that enter the volume $V$ with energy $E<0$ within a time interval, $t_0,t_0+\Delta t$, where $\Delta t$ can be arbitrarily small. 
To this aim, it is useful to define the {\it entrapment rate} as 
\begin{align}\label{eq:ratea1}
  C_{\rm trap}\equiv \lim_{\Delta t\to 0}\frac{\Delta N_b}{\Delta t}.
\end{align}
The chief difficulty in deriving $\Delta N_b=N_b(t)-N_b(t_0)$ from Equation~(\ref{eq:Nb}) lies in the statistical correlation between the location of particles at two consecutive times, $t=t_0$ and $t_0+\Delta t$. Indeed, it is easy to see that if the time interval becomes small, $\Delta t\to 0$, one should expect a strong correlation between the values of $N_b$ measured at the two instants of time because many particles found within the volume $V$ at $t_0$ still remain in this region at $t_0+\Delta t$. Such correlations vanish when the length of $\Delta t$ is larger than the average time that a particle spends in the vicinity of $m_\star$. In this case, particles identified at $t_0$ leave the region under observation and are replaced by a new set of particles at $t_0+\Delta t$. Hence, consecutive measurements of $N_b$ become statistically independent, and the problem can be treated within a Markovian framework.

Here, the theory of Smoluchowski (1916) is used to compute the entrapment rate~(\ref{eq:ratea1}) between arbitrarily-close time intervals. This theory was originally devised to model the fluctuations of colloid concentrations in a liquid, and later used by Chandrasekhar (1941, 1943) in the context of self-gravitating systems. The general idea rests on the notion of ``probability after-effect'', or {\it Wahrscheinlichkeitsnachwirkung}, which describes the statistics of correlated observations at different instants of time. In this paper, we use the results of Smoluchowski (1916) without giving formal proof (interested readers are directed to \S III of Chandrasekhar 1943 for a detailed account of the theory). The main assumptions are (1) that the motions of individual particles are independent of each other, and (2) that all positions within the volume $V$ have the same {\it a priori} probability to be sampled from an homogeneous distribution. Under these conditions, Smoluchowski (1916) shows that the probability $P_N(t)$ that at some later time there are still $N$ particles inside $V$ follows a law of decay that is analogous to the law of decay of radioactive substances
\begin{align}\label{eq:PN}
P_N(t)\d t =e^{-t/T} \d t/T,
\end{align}
with $T$ being the mean-life of a state in which the number of particles within the volume $V$ remains constant. Under the local approximation, $r\ll d$, the fluctuation mean-life roughly corresponds to the time that a particle moving on a straight-line trajectory takes to cross the size of the sphere
\begin{align}\label{eq:Ta}
  T(r)=\sqrt\frac{2\pi}{3}\frac{r}{\langle v^2\rangle^{1/2}},
\end{align}
where $\langle v^2\rangle^{1/2}$ is the average relative velocity between $m_\star$ and the background particles, which can be estimated from~(\ref{eq:v2a}) under the Maxwellian approximation. Note that the time-scale~(\ref{eq:Ta}) roughly corresponds to the time that a field particle moving on a straight line with a velocity $\langle v^2\rangle^{1/2}$ takes to cross the volume $V$.

Smoluchowski (1916) then shows that the probability that $N_e$ particles enter the volume $V$ at any arbitrary time between $t_0, t_0+\Delta t$ follows a Poisson distribution
\begin{align}\label{eq:Ne}
  W(N_e)=\frac{e^{-NP}(N P)^{N_e}}{N_e!},
\end{align}
where $N=n V$ is the average number of particles in $V$, and $P$ is the so-called {\it probability after-effect factor}
\begin{align}\label{eq:P}
  P = \frac{\Delta t}{T},
\end{align}
which represents the probability that a particle initially inside a given volume emerges from $V$ within the time-interval $\Delta t$ (see also Chandrasekhar 1943, p.53). As expected, when the time-interval is much shorter than the fluctuation mean-life particles do not have sufficient time to exit this region, hence the probability after-effect approaches $P\to 0$ in the limit $\Delta t/T\to 0$. 
Direct inspection of the Poisson distribution~(\ref{eq:Ne}) shows that the average number of particles leaving the volume is $N_l=N P=N\Delta t/T$. In dynamical equilibrium, this is equal to the number of entering particles, hence $N_e=N_l=N\Delta t/T$. 

Under the assumption that particles within the volume $V=4\pi r^3/3$ are statistically uncorrelated, the above arguments must hold regardless their binding energy to the point-mass $m_\star$. 

The entrapment rate~({\ref{eq:ratea1}) at which particles with negative binding energies enter the volume $V$ then follows from Equations~(\ref{eq:Ne}) and~(\ref{eq:P}), where $N$ is replaced by $N_b$. Inserting~(\ref{eq:Nb}) and~(\ref{eq:Ta}) yields 
\begin{align}\label{eq:ratea}
  C_{\rm trap}(r)&=\lim_{\Delta t\to 0}\,N_b\frac{P}{\Delta t}=\frac{N_b}{T}\\\nonumber
   &=  \frac{16}{3}\bigg(\frac{2}{3}\bigg)^{1/2}\sqrt{\langle v^2\rangle}(Gm_\star)^{3/2}\frac{n}{\sigma^3}e^{-V_\star^2/(2\sigma^2)}r^{1/2}.
\end{align}

Time-integrating~(\ref{eq:ratea}) returns a number of entrapment events that grows linearly with time. 
\begin{align}\label{eq:Nat}
  N_{\rm trap}(r,t)&=\int_0^t\d t\,C_{\rm trap}= N_b\,\frac{t}{T}\\ \nonumber
  &=\frac{16}{3}\bigg(\frac{2}{3}\bigg)^{1/2}\sqrt{\langle v^2\rangle}(Gm_\star)^{3/2}\frac{n}{\sigma^3}e^{-V_\star^2/(2\sigma^2)}r^{1/2}\,t.
\end{align}
Similarly, it is useful to compute the number of field particles leaving or entering the volume $V$ in the time interval $t$
\begin{align}\label{eq:Net}
  N_e(r,t)&=N_l(r,t)=N\frac{t}{T} \\ \nonumber
  &=\bigg(\frac{8\pi}{3}\bigg)^{1/2}\sqrt{\langle v^2\rangle}\,n\,r^2\,t,
\end{align}
with $N=n V$.

Notice that~(\ref{eq:Nat}) and~(\ref{eq:Net}) cross each other at a radius comparable to the critical radius, $r_0=2Gm_\star/\sigma^2$. More precisely, solving $N_{\rm trap}(r_\epsilon,t)=N_e(r_\epsilon,t)$ yields 
\begin{align}\label{eq:r_eps}
  r_\epsilon=\bigg(\frac{16}{9\pi}\bigg)^{1/3}e^{-V_\star^2/(3\sigma^2)}\frac{Gm_\star}{\sigma^2} \simeq 0.4\, e^{-V_\star^2/(3\sigma^2)}r_0.
\end{align}
Below this ``thermal'' critical radius\footnote{Here, the word ``thermal'' highlights the Maxwellian factor $\exp[-V_\star^2/(3\sigma^2_\star)]$ multiplying the critical radius.} the theory is not valid, as it predicts that the number of newly trapped particles exceeds the total number of particles entering the volume, $N_{\rm trap}>N_e$ at $r<r_\epsilon$, a result with no physical meaning within our idealized theoretical framework. This scale becomes arbitrarily small as the attractor mass vanishes ($m_\star\to 0$), or its speed increases ($V_\star\to \infty$).

The thermal critical radius $r_\epsilon$ defines the {\it sphere of influence} of the point-mass $m_\star$, such that the dynamics of field particles found within the volume $V_\epsilon=4\pi r^3_\epsilon/3$ are completely dominated by the Keplerian potential, $\Phi_\star$. Indeed, comparison of the mean kinetic energy of field particles $\mathcal{T}=3\sigma^2/2$ and the potential energy at $r=r_\epsilon$ derived from Equation~(\ref{eq:r_eps}) yields
 \begin{align}\label{eq:phi_eps}
  \Phi_\star(r_\epsilon)&=-\frac{Gm_\star}{r_\epsilon} \\
  &=-\frac{1}{2}\bigg(\frac{9\pi}{2}\bigg)^{1/3}e^{V_\star/(3\sigma^2)}\,\sigma^2\approx -0.81\,e^{V_\star/(3\sigma^2)}\,\mathcal{T}.
  \end{align}
 Hence, particles located within a distance $r\lesssim r_\epsilon$ have binding energies that exceed the mean kinetic energy of the field, $|\Phi_\star|\gtrsim {\mathcal{T}}$. In general, the probability to find {\it unbound interlopers} crossing the volume $V_\epsilon$ is small, and it decreases further for point-masses that are not at rest with the galactic background ($V_\star\ne 0$).

\subsection{Survival}\label{sec:survival}
As outlined in the Introduction, tidal forces can both decelerate and accelerate interstellar particles in the proximity of the point-mass $m_\star$. This means that the same dynamical mechanism that allows interstellar bodies to lose energy will also allow them to escape from the Sun's gravitational field. Following Napier et al. (2021b), it is useful to define a {\it dynamical lifetime function}, $f_{\rm surv}(t)$, which determines the fraction of captured objects that remain in bound orbits as a function of time since entrapment occurred. Time-integrating~(\ref{eq:ratea}) yields
\begin{align}\label{eq:Nsurv}
  N_{\rm surv}(r,t)=\int_0^t\d t\,f_{\rm surv}(t)\,C_{\rm trap}=N_b \,\frac{T_s}{T}\alpha(t),
\end{align}
where $T_s$ is a characteristic survival time, and $\alpha(t)$ is a dimensionless {\it abundance parameter} defined as
\begin{align}\label{eq:alpha}
 \alpha(t)\equiv \frac{1}{T_s}\int_0^t\d t\,f_{\rm surv}(t).
\end{align}
Numerical tests in Section~\ref{sec:tests} show that Equation~(\ref{eq:alpha}) converges asymptotically to a value of order unity, $\alpha(t)\to \alpha$ at $t\to \infty$, such that the number of surviving particles reaches a {\it steady-state} value
\begin{align}\label{eq:Nss}
   N_{\rm surv}(r,t)\to N_{\rm ss}(r)=N_b \,\frac{T_s}{T}\alpha~~~~~~~{\rm for}~~~t/T_s\to \infty.
\end{align}
Equation~(\ref{eq:Nss}) exhibits three different limiting cases depending on the type of trajectories on which tidally-trapped objects move:
\begin{enumerate}
  \item Unstable orbits. Binding energies only become negative during the duration of a single encounter. The survival time in this case is comparable to the average time that a random body spends within the volume $V$, i.e. $T_s\approx T \approx r/\sqrt{\langle v^2\rangle}$. Setting $T_s=T$ in Equation~(\ref{eq:Nss}) returns an asymptotic, {\it steady-state} value $N_{\rm ss}=N_b\,\alpha$ for $t\gg T_s$.
  \item  Semi-stable orbits. Particles remain bound during several orbital revolutions before being released back to the galactic potential. The survival time is long but finite, $T_s\gg T$. Hence, the number of bound objects converges to $N_{\rm ss}\approx N_b (T_s/T)\alpha \gg N_b$ on a time-scale $t\gg T_s$.
      \item Stable orbits. Trapped objects remain bound for an arbitrarily long interval of time. The survival time diverges $T_s/T\to \infty$, and the survival fraction approaches $f_{\rm surv}\to 1$. Comparison of~(\ref{eq:Nat}) and~(\ref{eq:Nsurv}) yields $N_{\rm surv}(t)=N_{\rm trap}(t)=N_b\,(t/T_s)$, which grows linearly with time. As a result, no steady-state exists in this limit.
  \end{enumerate}

In the collisionless approximation, interstellar particles cross the volume $V$  on statistically uncorrelated trajectories. Under this condition, the amount of time that a random field particle spends in the volume $V$ is governed by Smoluchowski (1916) probability function~(\ref{eq:PN}). Accordingly, the dynamical lifetime is expected to decay exponentially as $f_{\rm surv}(t)=\exp(-t/T)$, where $T$ is the fluctuation mean-life~(\ref{eq:Ta}). The characteristic survival time that a random bound particle spends in the volume $V$ is therefore comparable to the duration a fly-by encounter, $T_s=T$, which means that tidally-trapped objects move on unstable orbits in the Keplerian potential $\Phi_\star$. 
Inserting an exponential-decaying lifetime function in Equation~(\ref{eq:alpha}) and setting $T_s=T$ yields an abundance parameter
\begin{align}\label{eq:alpha_less}
 \alpha=\lim_{t\to \infty}  \frac{1}{T}\int_0^t\d t\,\exp(-t/T)=1.
\end{align}
Hence, the steady-state number of bound particles around the point-mass $m_\star$ can be derived from~(\ref{eq:Nb}) and~(\ref{eq:Nss}) by setting $T_s=T$ and $\alpha=1$, which yields
\begin{align}\label{eq:Nss_asym}
  N_{\rm ss}(r)&= N_b(r) \\ \nonumber
  &=\frac{32\sqrt{\pi}}{9}(G m_\star)^{3/2}\frac{n}{\sigma^3}e^{-V_\star^2/(2\sigma^2)}r^{3/2}~~~~~~~{\rm for}~~~~t\gg T.
\end{align}

As shown in \S\ref{sec:acc}, these predictions fail on scales comparable or smaller than the thermal critical radius, where Equation~(\ref{eq:Nss_asym}) predicts a number of bound particles that exceeds the number of particles enclosed in the volume element $V$. More precisely, solving $N_{\rm ss}(r'_\epsilon)=N(r'_\epsilon)$ and inserting~(\ref{eq:r_eps}) yields $r'_\epsilon=2^{2/3} r_\epsilon$. Therefore, Equation~(\ref{eq:Nss_asym}) should not be applied on scales below the thermal critical radius $r\lesssim r_\epsilon$.
Section~\ref{sec:tests} tests the above theoretical expectations with live $N$-body models.

\subsection{Steady-state profiles}\label{sec:profile}
The presence of trapped interstellar bodies around the central star leads to the formation of an extended `halo' that approaches an equilibrium state as the number of particles being trapped equals the number being tidally stripped.

In the collisionless limit, the steady-state density enhancement due to the presence of bound interstellar material can be directly derived from~(\ref{eq:Nss_asym}) as
\begin{align}\label{eq:delta}
  \delta(r)&\equiv \frac{1}{4\pi\,n\,r^2}\frac{\d N_{\rm ss}}{\d r} \\ \nonumber
  &=\frac{4}{3\sqrt{\pi}}\frac{(Gm_\star)^{3/2}}{\sigma^3} e^{-V_\star^2/(2\sigma^2)}\frac{1}{r^{3/2}},
\end{align}
which scales as $\delta \sim (v_c/\sigma)^3 \sim r^{-3/2}$, where $v_c(r)=\sqrt{Gm_\star/r}$ is the circular velocity at a distance $r$ from the Keplerian potential. This profile -- generally known as a `density spike'-- was originally found by Gondolo \& Silk (1999) by analyzing the spatial distribution of particles with a constant phase-space density around a Black Hole in isolation.

Of particular relevance is the distance at which the density of trapped particles equals the background interstellar density. Using~(\ref{eq:delta}), it is straightforward to show that this happens at the thermal critical radius, $\delta(r_\epsilon)=1$, which calls for caution when extrapolating Equation~(\ref{eq:delta}) at $r\lesssim r_\epsilon$ (or $\delta\gtrsim 1$). We will come back to this issue in Section~\ref{sec:tests}.

The velocity dispersion of the bound halo is found from~(\ref{eq:delta}) by solving the isotropic Jeans equations 
\begin{align}\label{eq:vdisp}
  \sigma_h^2(r)&=\frac{1}{\delta(r)}\int_r^\infty\d r' \delta(r')\bigg|\frac{\d\Phi_\star}{\d r}\bigg| \\ \nonumber
  &=\frac{2}{5}\frac{Gm_\star}{r},
  \end{align}
which corresponds to a constant fraction of the circular velocity, $\sigma_h/v_c=(2/5)^{1/2}\approx 0.632$ at all radii.

It is straightforward to show that the mean phase-space density of halo particles is constant across the volume $V$. Combination of~(\ref{eq:delta}) and~(\ref{eq:vdisp}) yields
\begin{align}\label{eq:Qacc}
  Q_h&\equiv \frac{n\,\delta(r)}{\sigma^3_h(r)}\\\nonumber
  &=\frac{5}{3}\bigg(\frac{10}{\pi}\bigg)^{1/2}e^{-V_\star^2/(2\sigma^2)}\,Q_g,
\end{align}
where $Q_g=n/\sigma^3$ is the field phase-space density. Interestingly, the ratio $Q_h/Q_g$ solely depends on the speed of the attractor relative to the host galaxy, $V_\star/\sigma$.
Interstellar particles bound to a point-mass at rest have higher phase-space densities than those in the Galactic background, $Q_h(V_\star=0)\simeq 2.97 \,Q_g$. The halo mean phase-space density drops for moving objects, such that $Q_h<Q_g$ for objects travelling with a speed $V_\star> 1.476\,\sigma$.

\subsection{Orbits in isolation}\label{sec:orbits}
Up to this point, we have not made any explicit assumption on the motion of particles trapped in the halo. This Section derives the equilibrium distribution of a large ensemble of these objects orbiting around a point-mass in isolation, such that the only force acting on bound particles corresponds to the gravitational attraction of $m_\star$.
Invoking isolation is a useful theoretical {\it fudge} that allows us to compute ensemble-averaged properties of trapped particles as a superposition of orbits that conserve specific energy $E=v^2/2 - Gm_\star/r$ and angular momentum $\bb L=\bb r\times \bb v$, making the problem mathematically tractable.
However, one should bear in mind that this assumption is particularly poor for weakly-bound particles, as illustrated in Fig.~\ref{fig:e}.

Following Equation~(\ref{eq:Qacc}), we assume that energetically-bound particles are distributed homogeneously in phase-space on distance scales $r\ll d$, with a distribution function that is approximately constant across the volume $V=4\pi r^3/3$, i.e. $f({\bb r},{\bb v})= f_0$. The spectrum of semi-major axes and eccentricities is found by mapping points in integral-of-motion space onto the phase-space volume $\d^6 \Omega = \d^3 r\,\d^3 v$ by means of the Jacobian $\omega(E,L)=|\partial^6 \Omega/\partial E\partial L|=8 \pi^2 L\, P$, also known as {\it density of states} (e.g. see Appendix A of P15), where $P=2\pi a^{3/2}/\sqrt{G m_\star}$ is the orbital period and $L=[Gm_\star\,a(1-e^2)]^{1/2}$ is the angular momentum of particles with a given energy. Writing $\d E=G m_\star \d a/(2a^2)$ and $L\, \d L=- G m_\star \,a\, e \d e$ yields $\d^6 \Omega=8 \pi^2 L\, P\, \d E\d L=8\pi^3(Gm_\star)^{3/2}\,e\,a^{1/2}\, \d(-e)\d a$. Hence, 
\begin{align}\label{eq:omega_DF}
  \omega(a,e)=8\pi^3(Gm_\star)^{3/2}\,e\,a^{1/2},
\end{align}
which recovers Equation~(19) of Dehnen et al. (2021).
The probability to find particles with semi-major axes and eccentricities in the intervals $(a,a+\d a)$ and $(e,e+\d e)$ is
\begin{align}\label{eq:DF}
  p(a,e)= \omega(a,e)\,f(a,e) = 8\pi^3(Gm_\star)^{3/2}\,e\,a^{1/2}\,f_0.
\end{align}
Equation~(\ref{eq:DF}) reveals a few points of interest. Notice first that the distribution of integrals has a separable form, $p(a,e)=p(a)p(e)$, with a semi-major axis distribution that follows a power-law
\begin{align}\label{eq:pa}
  p(a)\d a\sim a^{1/2}\d a.
\end{align}
The fact that $p(a)\to 0$ in the limit $a\to 0$ highlights the low probability to trap particles onto tightly-bound orbits by chance. The eccentricity distribution in Equation~(\ref{eq:DF}) is `thermal'
\begin{align}\label{eq:pecc}
  p(e)\d e=2e\,\d e,
\end{align}
which indicates that trapped particles tend to fall in on eccentric orbits.
As a sanity check, Appendix A shows that the number density profile associated with the distribution function~(\ref{eq:DF}) recovers Equation~(\ref{eq:delta}) derived from Smoluchowski's (1916) statistical theory. Appendix B proves that Equations~(\ref{eq:DF}) also describes the orbital distribution of wide binaries that form via chance entrapment of uncorrelated stars with Maxwellian velocities.

The closest approach of a trapped object to the sun (i.e. the orbital pericentre) is related to the semi-major axis and eccentricity as $r_p=a(1-e)$. The fraction of objects within a pericentre interval $r_p,r_p+\d r_p$ can be straightforwardly calculated from~(\ref{eq:pecc}) as (c.f. eqs 3 and 4 of Hills 1981)
\begin{align}\label{eq:prperi}
  p(r_p)\,\d r_p=\frac{2}{a}\bigg(1-\frac{r_p}{a}\bigg)\,\d r_p.
\end{align}
Notice that for $r_p\ll a$ the distribution $p(r_p)$ becomes independent of $r_p$, which therefore implies that trapped halo particles reaching the inner solar system have a homogeneous distribution of perihelia. This has important implications for the detection of trapped interstellar objects from Earth, as discussed in Section~\ref{sec:discussion}.

\section{Numerical tests}\label{sec:tests}
Given the heuristic nature of the theory presented in \S\ref{sec:stats}, it is useful to run controlled $N$-body tests that {\it falsify} two key assumptions on which the theory rests. Namely, we use galaxy models where (i) the background density is not homogeneous, and (ii) the velocity function is not Maxwellian. In these experiments, the motion a point-mass $m_\star$ (the ``Sun'') is followed as it travels on a circular orbit through a self-gravitating Dehnen (1993) sphere in dynamical equilibrium (the ``DM halo''). At each time-step, we identify particles that become gravitationally bound to the point mass (i.e. binding energy flips from positive to negative), compute their orbital elements in the Keplerian potential $\Phi_\star$, and count the number of particles stripped by Galactic tides (i.e. binding energy flips from negative to positive).

To compute the orbits of the point-mass $m_\star$ and halo particles in the host galaxy, we solve the equations of motion~(\ref{eq:eqmotsun}) and~(\ref{eq:eqmot}), respectively, using a leap-frog technique with varying time-steps (e.g. Press 1992). Details on the numerical integration are given in Appendix~\ref{sec:con}.

In addition, for a better understanding of the effect of $m_\star$ on the local galactic background we run $N$-body experiments where the gravitational force induced by the point-mass is removed from the equations of motion ($F_\star=0$). Hence, in these so-called ``collisionless'' experiments field particles follow orbits in a smooth galaxy potential $\Phi_g$ and do not feel the gravitational attraction of the point-mass, $m_\star$. Their phase-space distribution obeys the collisionless Boltzmann equation, which provides a closer match to the idealized conditions on which the statistical theory presented in \S\ref{sec:stats} rests.
 
\subsection{Set up}\label{sec:setup}
The first step of this Section is to generate $N$-body realizations of a Dehnen (1993) sphere in dynamical equilibrium. The gravitational potential of these objects has an analytical form
\begin{align}\label{eq:phi_dehn}
\Phi_g(R)=\frac{4\pi G\rho_0}{3-\gamma}\times\begin{cases}
-\frac{1}{2-\gamma}\big[1-\big(\frac{R}{R+R_0}\big)^{2-\gamma}\big] & ,\gamma\ne 2 \\ 
\ln\big(\frac{R}{R+R_0}\big) & , \gamma=2.
\end{cases}
\end{align}
where $R$ is the galactocentric radius. The total mass is $M_g=4\pi \rho_0 R_0^3/(3-\gamma)$, and the number density profile is
\begin{align}\label{eq:n_dehn}
n(R)=\frac{n_0}{(R/R_0)^\gamma[1+(R/R_0)^{4-\gamma}]},
\end{align}
with $n_0$ chosen such that $4\pi\int_0^\infty\d r\, r^2\, n(r)=N_g$. The density scale-length can be directly derived from~(\ref{eq:n_dehn}) 
\begin{align}\label{eq:d_dehn}
  d(R)=\bigg|\frac{\nabla n}{n}\bigg|^{-1}_R=R\frac{R+R_0}{4R+\gamma R_0}.
\end{align}
Recall that at a given galactocentric radius, the local approximation is valid on distance scales $r\ll d$, where $r$ is measured relative to the point-mass location.

To assign orbital velocities in a way that guarantees dynamical equilibrium, the distribution function is calculated using Eddington (1916) inversion
\begin{align}\label{eq:eddin}
  f(E)=\frac{1}{\sqrt{8}\pi^2}\bigg[\int_E^0\frac{\d\Phi}{\sqrt{\Phi-E}}\frac{\d^2 n}{\d\Phi^2}+\frac{1}{\sqrt{-E}}\bigg(\frac{\d n}{\d \Phi}\bigg)_{\Phi=0}\bigg],
\end{align}
where $n[R(\Phi_g)]$ corresponds to the profile~(\ref{eq:n_dehn}) expressed as a function of the potential~(\ref{eq:phi_dehn}). Note that the velocity distribution resulting from Eddington's equation~(\ref{eq:eddin}) is {\it not} Maxwellian. To generate halo models in equilibrium, positions and velocities for $N_g=5\times 10^7$ particles are drawn randomly from the distribution function~(\ref{eq:eddin}), with unity vectors isotropically distributed over the surface of a sphere.

To guide the physical interpretation of our results, we choose a dark matter halo with a mass and scale radius that match those of the Aquarius simulation, $M_g=1.84\times 10^{12}M_\odot$, $R_0=15.3\kpc$ and $\gamma=1$ (Springel et al. 2008). Subsequently, we place point-mass attractors at a fixed distance from the halo centre, $R_\star=0.065\,R_s$, with a circular velocity $V_\star=V_c(R_\star)=172.4\kms$. The density scale-length~(\ref{eq:d_dehn}) at this galactocentric radius is $d(R_\star)=0.84\kpc$, and the local velocity dispersion is $\sigma(R_\star)=196\kms$. Under the Maxwellian approximation, the average relative velocity between the point-mass and the surrounding DM particles is therefore $\sqrt{\langle v^2\rangle}=\sqrt{V_\star^2+3\sigma^2}\simeq 381\kms$. The fluctuation mean-life~(\ref{eq:Ta}) at a relative distance $r=d$ is $T(d)=3.12 \myr$, whereas the mean phase-space density of trapped halo particles predicted by Equation~(\ref{eq:Qacc}) is $Q_h/Q_g\approx 2$.

For illustration, we consider three values for $m_\star$ that are specifically chosen to test various aspects of the theory presented in \S\ref{sec:stats}.
Recall that the theoretical model works on distance scales where the local approximation is valid, $r\ll d$, and the escape velocity is much smaller than the velocity dispersion of the field, $v_{\rm esc}(r)/\sigma \ll 1$. With this in mind, we choose $m_\star/\msol=10^8, 3\times 10^8$ and $10^9$, which have characteristic escape velocities 
$v_{\rm esc}(d)=(2Gm_\star/d)^{1/2}= \{0.16, 0.28,0.51\}\,\sigma$, respectively. We shall see below that the lightest model mostly evolves in a collisionless regime, while the most massive one shows noticeable deviations from a collisionless behaviour.

Finally, the integration time ($t_f$) is chosen to obtain a statistically-meaningful number of trapping events. In particular, setting $t_f/T(d)=100$ in Equation~(\ref{eq:Nat}) yields $N_{\rm trap}(d,t_f)\sim \{1.1, 57.3, 348.6\} \times 10^4$ trapped particles for $m_\star/\msol=10^8, 3\times 10^8$ and $10^9$, respectively.

Notice that the $N$-body set-up is still far from reaching the resolution required to model stellar-size objects. Unfortunately, given that the number of bound particles~(\ref{eq:Nss_asym}) drops as $N_{\rm trap}\sim m_\star^{3/2}$, it becomes numerically demanding to run experiments with lighter point-mass attractors. For example, in order to find $N_{\rm trap}\sim 10^2$ objects bound to a stellar-size object ($m_\star/M_g\sim 10^{-12}$) one would need to increase the number of halo particles by $\sim 12$ orders of magnitude (i.e. $N_g\sim 10^{20}$), which lies far beyond current computational capabilities.

 \subsection{Results}\label{sec:results}

 \begin{figure*}
\begin{center}
\includegraphics[width=176mm]{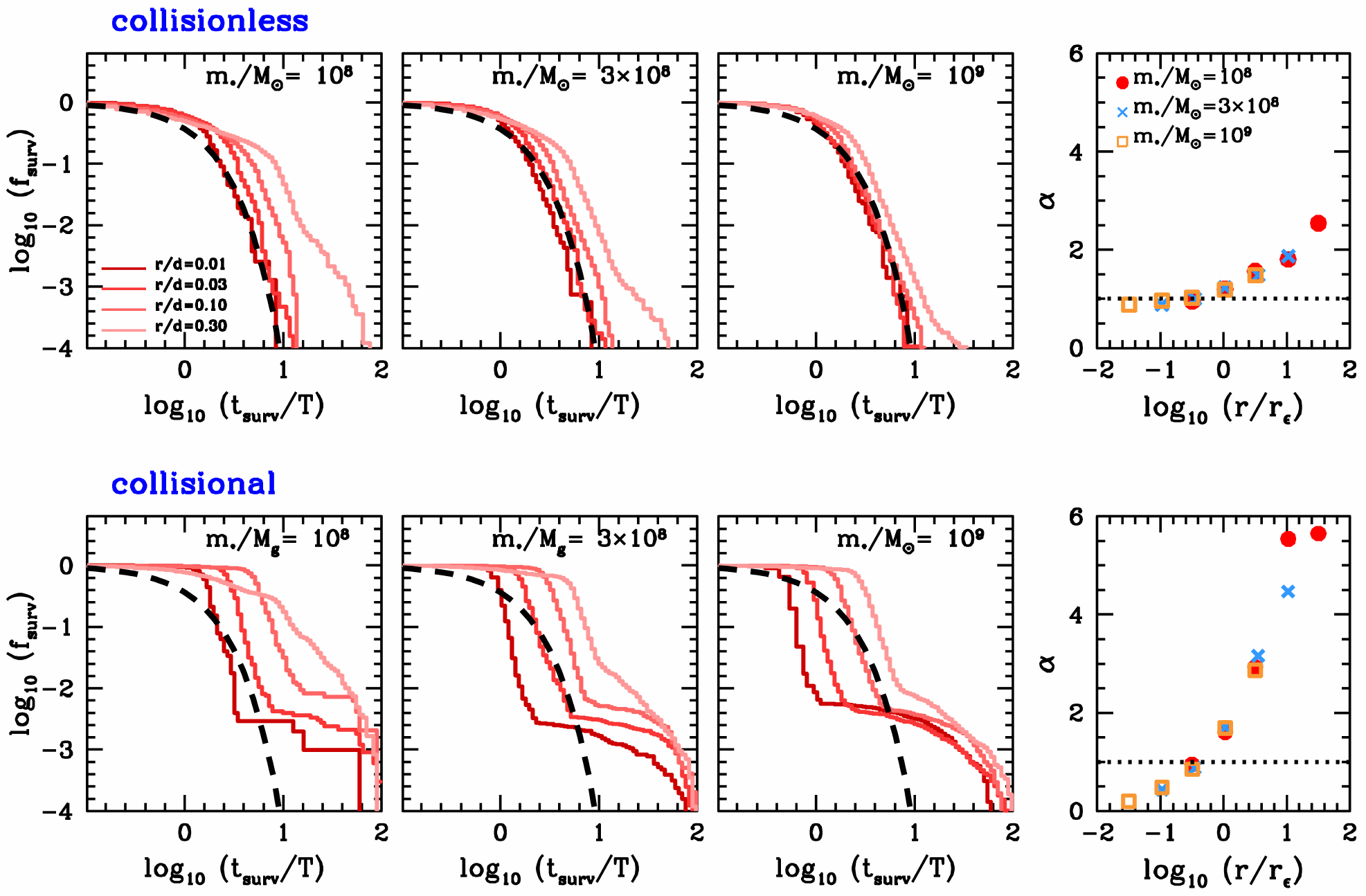}
\end{center}
\caption{Distribution of survival times, $t_{\rm surv}$, for collisionless and collisional models (top and bottom panels, respectively) measured in units of the fluctuation mean-life, $T$, Equation~(\ref{eq:Ta}). Here, $t_{\rm surv}$ is defined as the time interval over which a particle continuously has $E<0$ within the volume element $V=4\pi r^3/3$ centred at a point-mass $m_\star$. Lines are colour-coded according to volume radius ($r$) given in units of the local density scale-length $(d)$. Black-dashed lines show an exponentially-decaying dynamical survival fraction~(\ref{eq:PN}) with a characteristic survival time set equal to the fluctuation mean-life, $f_{\rm surv}(t)=\exp(-t/T)$. Right panels plot the abundance parameter $\alpha$ derived from Equation~(\ref{eq:alpha}) as a function of the volume size ($r$) measured in units of the thermal critical radius ($r_\epsilon$), Equation~(\ref{eq:r_eps}). }
\label{fig:time}
 \end{figure*}
 \subsubsection{Survival time}\label{sec:surv}
 The theory presented in \S\ref{sec:acc} assumes that the gravitational force of $m_\star$ induces negligible perturbations on the trajectories of nearby particles, which is the so-called ``collisionless'' approximation. Under this condition, the amount of time that an interstellar object spends in the volume $V$ follows Smoluchowski (1916)'s exponentially-decaying function~(\ref{eq:PN}), $\exp(-t/T)$, where the time-scale $T$ is defined by Equation~(\ref{eq:Ta}) and roughly corresponds to the average time that a particle moving on a straight line with a relative velocity $\langle v^2\rangle^{1/2}$ takes to cross the volume $V$. Thus, in a collisionless framework the typical time-span that a particle remains energetically bound to $m_\star$ is comparable to the duration of a flyby encounter, whereas the probability to find bound particles with survival times $t_{\rm surv}\gg T$ drops exponentially. 

 To test these expectations, Fig.~\ref{fig:time} plots the dynamical lifetime function ($f_{\rm surv}$) of particles bound to three different point-masses $m_\star$. Top and bottom panels respectively correspond to ``collisionless'' models, in which the force induced by the point-mass is removed from the equations of motion~(\ref{eq:eqmot}) ($F_\star=0$), and to ``collisional'' models, which do account for the gravitational force of the point-mass ($F_\star\ne 0$).  Recall that $t_{\rm surv}$ is defined as the time over which a particle located within a volume element $V=4\pi r^3/3$ continuously has $E<0$.
 
 As expected, models that neglect the point-mass self-gravity lead to a distribution of survival times that converges towards Smoluchowski (1916) exponentially-decaying function, $\exp(-t/T)$ (black-dashed lines), in the limit $r/d\to 0$. Consequently, the abundance parameter~(\ref{eq:alpha}) plotted in the right-most panel asymptotically approach unity, i.e. $\alpha \to 1$ as $r/d\to 0$. Notice that varying the attractor $m_\star$ has a visible impact on how fast these models converge towards the collisionless limit, with more massive models converging faster than low mass ones. Measuring the volume size $r$ in units of the thermal critical radius $r_\epsilon$ defined in Equation~(\ref{eq:r_eps}) reveals the self-similar behaviour of $\alpha$, which is solely controlled by the ratio $r/r_\epsilon$, such that $\alpha\approx 1$ for $r/r_\epsilon\lesssim 1$. Following Equation~(\ref{eq:Nss_asym}), this implies that the steady-state number of bound particles in the volume element $V$ converges to the value derived from the unperturbed distribution function of the host, i.e. $N_{\rm ss}=N_b$ for $r/r_\epsilon\lesssim 1$.

 Comparison with the corresponding collisional models in the bottom panels reveals a number of interesting features.
 The first noticeable result is the emergence of a sizeable population of particles that remain bound to the Keplerian potential $\Phi_\star$ for time intervals that can be as long as the integration time of the models, $t_{\rm surv}\sim t_f=100\,T$.
 This translates into a distribution of survival times that departs from the exponential shape predicted by Smoluchowski (1916) and becomes more akin to a step function, in which most trapped particles move on unstable orbits that are lost to galactic tides soon after being entrapped, $t_{\rm surv}\lesssim T$, and a sizeable population of objects that move on semi-stable orbits with long survival times $t_{\rm surv}\gg T$ (c.f. panel c) in Fig.~\ref{fig:xyz}).  
 As discussed in \S\ref{sec:discussion}, long-live objects are particularly interesting for detection experiments in the solar system.

 Interestingly, the abundance parameters shown in the bottom right-most panel of Fig.~\ref{fig:time} also exhibit a self-similar behaviour when the distance is plotted in units of the thermal critical radius, but the radial dependence has noticeable differences with the values derived from collisionless simulations. In particular, we find that the gravitational attraction of the point-mass reduces the average survival time of trapped particles at small distances, such that $\alpha\ll 1$ at $r\ll d$. This in turn implies a steady-state number of bound particles $N_{\rm ss}\lesssim N_b$ at $r\lesssim r_\epsilon$. On the other hand, $\alpha$ exhibits systematically larger values than in collisionless models at distances $r\gtrsim r_\epsilon$, suggesting that the gravitational attraction of the point-mass may affect the overall spatial distribution of tidally-trapped particles. We inspect this issue in \S\ref{sec:profiles}.

 \begin{figure*}
\begin{center}
\includegraphics[width=164mm]{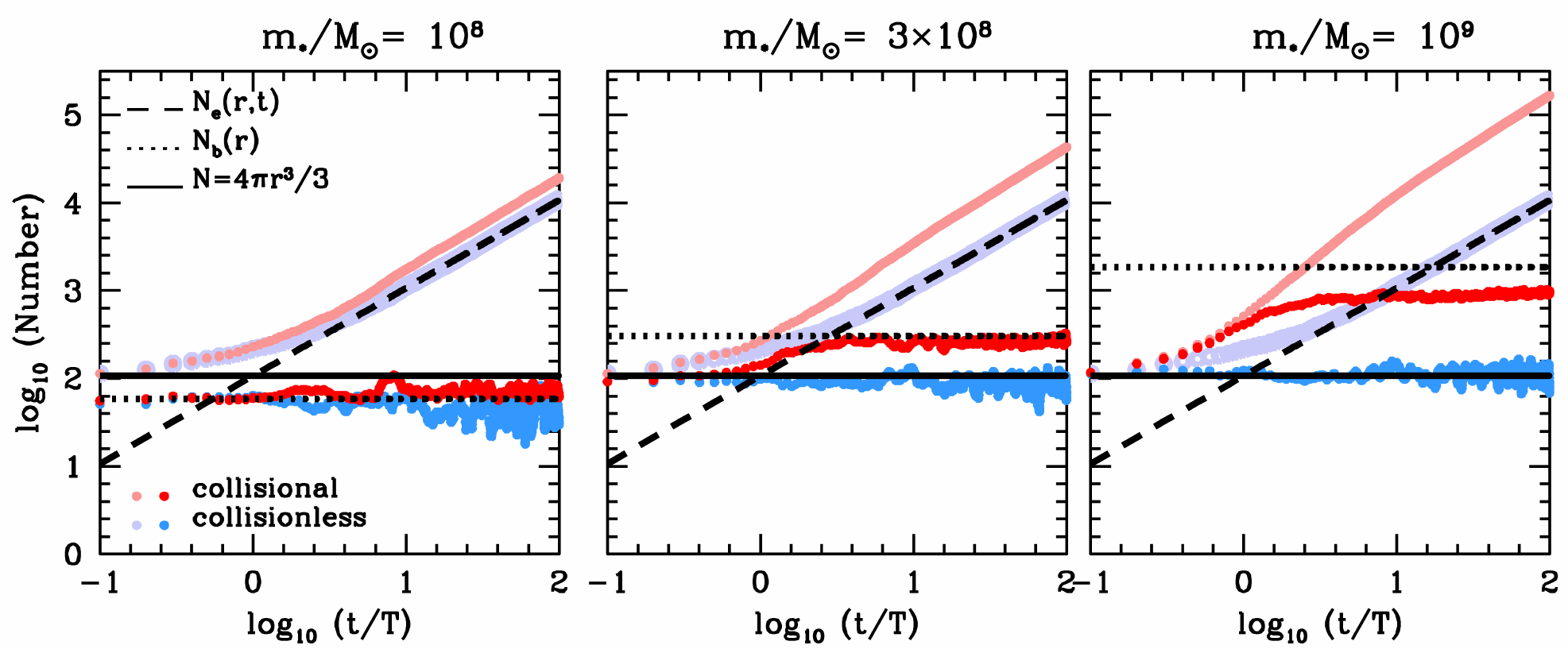}
\end{center}
\caption{ Time-evolution of the number of particles energetically-bound to three different point-masses $m_\star$ within the volume element $V=4\pi r^3/3$ for a fixed volume size $r=0.02\,d$. Time is given in units of the fluctuation mean-life, $T(r)$, Equation~(\ref{eq:Ta}). Red and blue symbols denote collisional and collisionless models, respectively.
  In collisionless models, the cumulative number of particles entering the volume $V$ (light-coloured symbols) grows linearly with time and is accurately described by Equation~(\ref{eq:Nat}) on time scales $t\gtrsim T$ (black-dashed lines) .
  The number of bound particles within $V$ (dark-coloured symbols) matches the steady-state value predicted by Equation~(\ref{eq:Nss_asym}) (dotted lines).
 In collisional models, the number of particle entering the volume $V$ also shows a linear growth with time, but with values that lie systematically above the relation~(\ref{eq:Nat}). The discrepancy increases systematically for more massive point-masses $m_\star$. Similarly, the population of bound particles is larger than expected from Equation~(\ref{eq:Nss_asym}). Note that in the middle and right panels, the number of bound particles that exceeds the {\it total} number of random field particles expected to be found within $V$, i.e. $N_b\gtrsim N=n\,V$. This effect is caused by the self-gravity of the point mass $m_\star$ and occurs in models with a volume size smaller than the thermal critical radius, i.e. $r\lesssim r_\epsilon$ (see text).
}
\label{fig:capt}
 \end{figure*}

 \begin{figure*}
\begin{center}
\includegraphics[width=164mm]{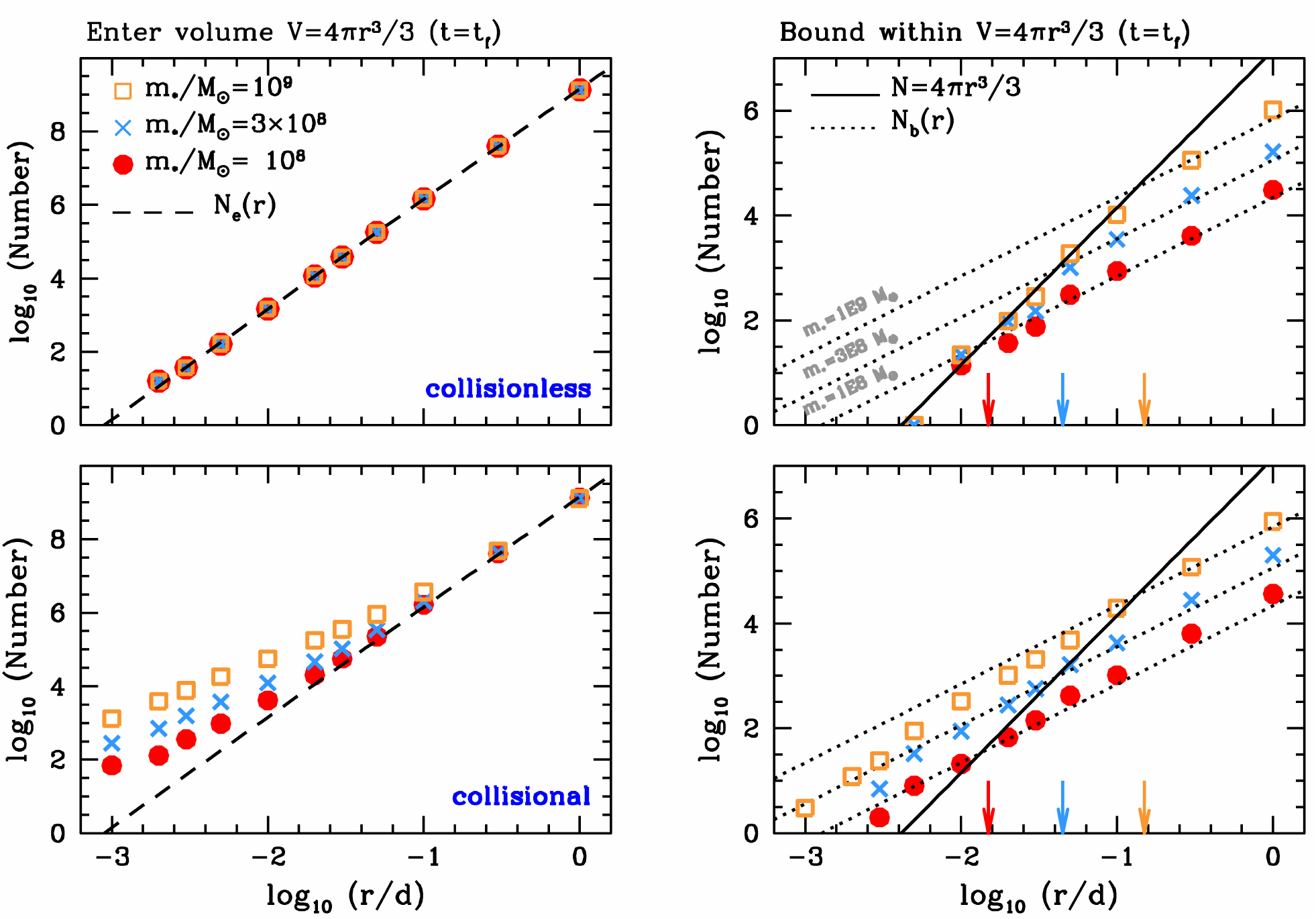}
\end{center}
\caption{{\it Left panels:} Time-evolution of the cumulative number of particles that enter the volume element $V=4\pi r^3/3$ at the time $t_f=100\,T$ as a function of volume size ($r$) normalized by the local density scale ($d$). Here, time is given in units of the fluctuation mean-life, $T$, Equation~(\ref{eq:Ta}). Upper and bottom panels correspond to collisionless ($F_\star=0$) and collisional ($F_\star\ne 0$) experiments, respectively. {\it Right panels:} Number of particles with negative binding energies at $t=t_f$ for the models shown in the left panels. Black-dotted lines correspond to the steady-state population of bound particles derived in a collisionless regime ($N_{\rm ss}$), Equation~(\ref{eq:Nss_asym}), whereas black-solid lines mark the relation $N=4\pi r^3/3$ expected in a constant-density background of field particles. Dotted and solid lines cross each other at at a distance $r=2^{2/3}\, r_\epsilon$, where $r_\epsilon$ is the thermal critical radius, Equation~(\ref{eq:r_eps}) (marked with vertical arrows for reference). 
  In collisionless models (upper panel), the number of bound particles cannot exceed the total number of particles in the volume element, $N_{\rm ss}\le N$. Most field particles entering the radius $r\lesssim r_{\rm eps}$ are energetically bound. Bottom panels show that the point-mass attraction removes this upper threshold in collisional models due to the presence of particles with long survival times $t_{\rm surv}\gg T$ (see Fig.~\ref{fig:time}). }
\label{fig:numt}
 \end{figure*}

  \subsubsection{Entrapment rates}\label{sec:rates}
  Next step is to check the analytical capture rates derived in \S\ref{sec:stats}. For illustration, we choose a relatively small radius ($r=0.02\,d$) centred around $m_\star$, and record the number of particles that enter \& leave the volume $V=4\pi r^3/3$ centred around $m_\star$, and monitor their binding energies to the Keplerian potential $\Phi_\star(r)=-G m_\star/r$ at each time step.
  
  Fig.~\ref{fig:capt} plots the time-evolution of the cumulative number of particles that enter the volume $V$ (light-coloured symbols), and the size of the population of particles with negative binding energies $E<0$ (dark-coloured symbols). Here, time measured in units of the fluctuation mean-life, $T(r=0.02d)$, Equation~(\ref{eq:Ta}). Red and blue symbols denote collisional and collisionless models, respectively.

  This Figure highlights a few important aspects of the problem. First, on sufficiently long time-scales $t\gtrsim T$ the cumulative number of particles crossing the volume $V$ grows linearly with time, which is to be expected from a constant entrapment rate. However, while collisionless models follow very closely the analytical expression~(\ref{eq:Nat}), $N_e=N\,t/T$ (black-dashed lines), in collisional simulations the number of field particles entering $V$ lies systematically above the theoretical expectations. The fact that the discrepancy grows in proportion to $m_\star$ indicates that the gravitational attraction of the point mass causes a systematic increase the local density of field particles, which in turn boosts the number of objects moving through $V$.
  
  Crucially, in all our models the number of bound particles within the volume $V$ converges to a steady-state value $N_{\rm surv}(t)\to N_{\rm ss}$ on time-scales $t\gtrsim T$. Here, the size of the bound population is measured as $N_{\rm surv}(t)=N_{\rm trap}(t)-N_{\rm unb}(t)$ (dark-coloured symbols), where $N_{\rm trap}(t)$ and $N_{\rm unb}(t)$ respectively are the cumulative number of particles trapped by and stripped from the Keplerian potential $\Phi_\star$ within a time interval $t$. For reference, solid lines mark the number of random particles enclosed with the volume element $V$ in a constant-density field, $N=n\,V$.
  
  For light point-mass models ($m_\star=10^8\msol$, left panel), the steady-state population of bound objects accurately matches the analytical expression~(\ref{eq:Nss_asym}), $N_{\rm surv}(t)\approx N_b$ (dotted-black lines), independently of whether or not the gravitational attraction of the point-mass is included in the equations of motion. However, as $m_\star$ increases, we start to observe noticeable differences between collisionless and collisional models. Namely, whereas in collisionless experiments the number of trapped particles is limited by the number of particles within a given volume element, such that $N_{\rm surv}(t)\le N$, collisional models do not obey this threshold. Instead, the steady-state number of bound particles lies between the number expected from a constant-density field and that predicted by Equation~(\ref{eq:Nss_asym}), i.e. $N\lesssim N_{\rm surv}(t)\lesssim N_b$. This result can be explained by the boosted rate of particles entering the volume $V$ in collisional models, and the fact that the abundance parameter falls below unity at small distances from the point-mass (see bottom-right panel in Fig.~\ref{fig:time}).
  
 For a better understanding of the above results, Fig.~\ref{fig:numt} shows how the total number of particles that enter the volume (left panels) and the size of the bound population (right panels) vary as a function of volume radius ($r$). As above, the we run models in both collisionless (upper panels) and collisional modes (bottom panels).

 The first noteworthy result in the upper-left panel is the excellent match between the number of particles entering the volume measured in the simulations and those predicted by Equation~(\ref{eq:Net}) (black-dashed line), $N_e(t_f)=N\,t_f/T$, over 9 orders of magnitude, which gives evidence that Smoluchowski's (1916) probability after-effect factor $T$ defined in Equation~(\ref{eq:P}) successfully accounts for the correlation in the number of particles measured at consecutive time intervals.
 Note also that collisionless models are insensitive to the value of $m_\star$ by construction. However, this symmetry is broken in collisional models plotted in the bottom-left panel, which show that the attractive force of the point-mass tends to increase the number of particles at small distances with respect to the values predicted by Equation~(\ref{eq:Net}), a discrepancy that grows as the value of $m_\star$ increases.
 
 Right panels show the number of bound particles found in the volume $V$ as a function of volume size for the three point-masses considered above. In collisionless models (upper-right panel), Equation~(\ref{eq:Nss_asym}) (black-dashed lines) slightly underestimates the numerical values at large radii from the point-mass. The source of the discrepancy can be traced back to the raising abundance parameter away from the point-mass, $\alpha\gtrsim 1$ at $r\gtrsim r_\epsilon$, shown in Fig.~\ref{fig:time}, whereas the analytical expression~(\ref{eq:Nss_asym}) adopts $\alpha=1$. However, the main mismatch is found at small volume sizes, where Equation~(\ref{eq:Nss_asym}) grossly overestimates the number of particles in the volume $V$ with $E<0$. This can be easily understood by noting that while the number of bound particles scales as $N_{\rm ss}\sim r^{3/2}$, the total number of random particles within a given volume element has a steeper radial dependence, $N\sim r^3$. The two curves approximately cross at the thermal critical radius, $r\approx r_\epsilon$, with $r_\epsilon$ given by Equation~(\ref{eq:r_eps}).  As discussed in \S\ref{sec:acc}, Equation~(\ref{eq:Nss_asym}) should not be extrapolated down to arbitrarily small volume sizes around $m_\star$.
 A simple way to account for this shortcoming is by modifying Equation~(\ref{eq:Nss_asym}) as follows
 \begin{align}\label{eq:Nss_reps}
  N_{\rm ss}(r)=
  \begin{cases}
    N& ,r\lesssim r_\epsilon \\
   N_b & , r_\epsilon\lesssim r\ll d,
  \end{cases}
 \end{align}
 which can be now applied at arbitrarily small distances from the point-mass. In what follows, it is worth bearing in mind that {\it all} field particles found within the thermal critical radius $r\ll r_\epsilon$ have negative binding energies. In other words, one should expect no interstellar visitors with $E>0$ at radii $r\ll r_\epsilon$.

 Adding the self-gravity of $m_\star$ to the equations of motion predominantly affects the distribution of bound particles in the vicinity of the point-mass. In particular, bottom-right panel shown that in collisional models the number of bound particles within a volume element $V$ can now exceed the number of random particles expected in a constant-density background, such that $N_{\rm ss}\gtrsim N=n\,V$ at $r\lesssim r_\epsilon$. This enhancement is due to the systematic increase in the number of particles entering the volume $V$, as shown in Fig.~\ref{fig:capt}.
 As a result, in collisional models the steady-state number of bound particles  lies systematically above the value found in collisionless experiments and below a naive extrapolation of Equation~(\ref{eq:Nss_asym}) below the critical thermal radius, such that $N\lesssim N_{\rm ss}\lesssim N_b$ at $r\lesssim r_\epsilon$, with $N_b$ given by Equation~(\ref{eq:Nb}) (shown with black-dotted lines).

    \begin{figure}
\begin{center}
\includegraphics[width=176mm]{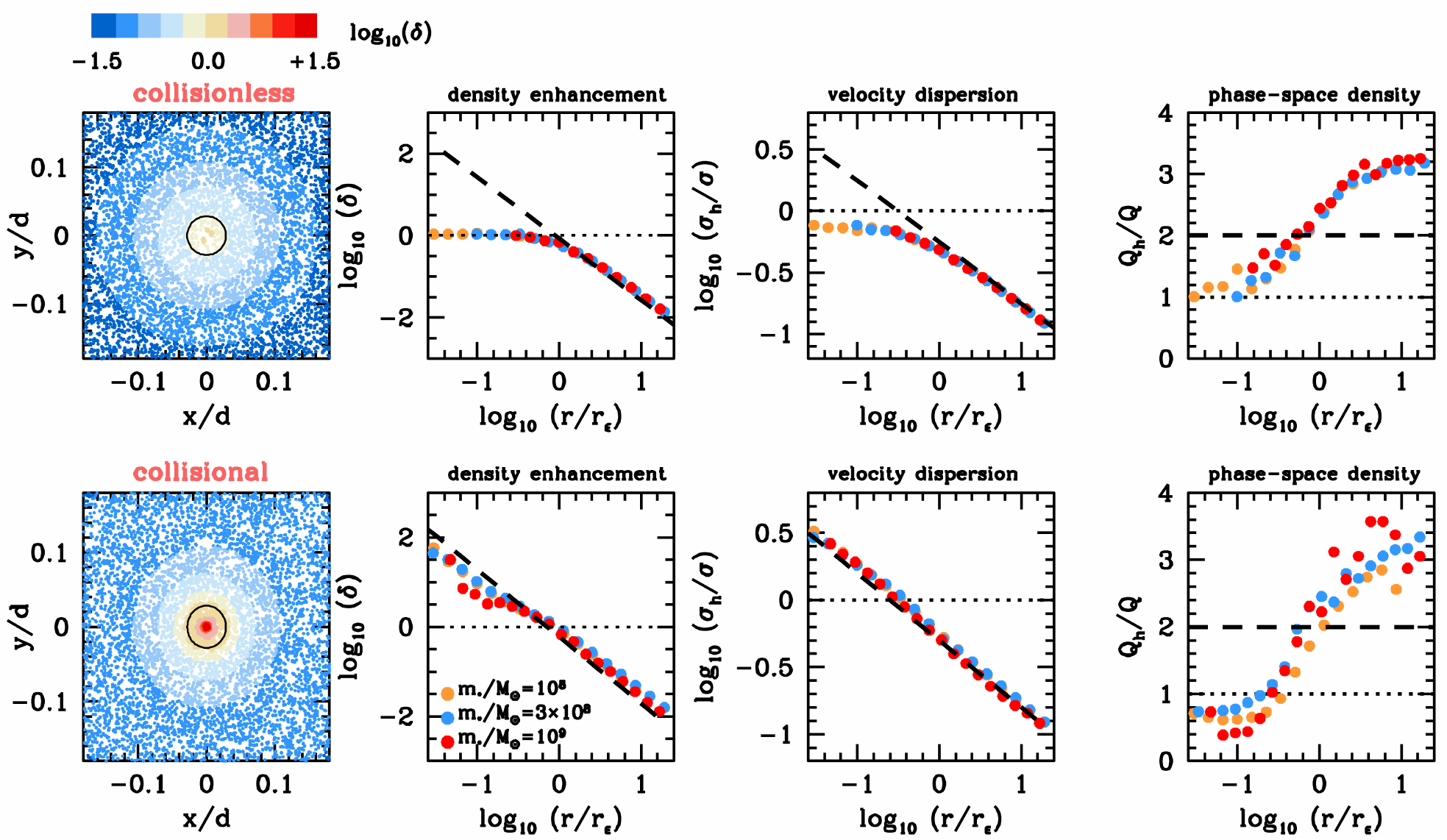}
\end{center}
\caption{{\it Left panels:} Spatial locations of halo particles bound to a point-mass $m_\star=3\times 10^8\msol$. Colours denote mean density enhancements measured at the location of each individual particle ($\delta$). For reference, the thermal critical radius ($r_\epsilon$) is marked with a black circle.
  {\it Middle-left panels:} density enhancement profile of halo particles, $\delta(r)$, as a function of distance from the central point-mass. Theoretical curves~(\ref{eq:delta}) are shown with black-dashed lines. Black-dotted particles denote the mean density of field particles ($\delta=1$).
  {\it Middle-right panels:} One-dimensional velocity dispersion profiles of bound halo particles as a function of distance from the point-mass. The theoretical profile~(\ref{eq:vdisp}) is shown with a black-dashed line.
{\it Right panels:} Mean phase-space density of bound halo particles ($Q_h=n\,\delta/\sigma^3_h$) as a function of distance. Black-dotted particles denote the phase-space density of the field ($Q_h/Q_g=1$).
}
\label{fig:prof_halo}
    \end{figure}
    
    \subsubsection{Profiles}\label{sec:profiles}
    Over time, tidally-trapped particles build up a`halo' around the point-mass $m_\star$.    
    Fig.~\ref{fig:prof_halo} shows the spatial and kinematic distribution of particles bound to the Keplerian potential $\Phi_\star$, together with the analytical profiles derived in \S\ref{sec:profile}. 
    Left panels plot the positions of bound particles on the orbital plane of $m_\star$ measured from the intermediate-mass model with $m_\star=3\times 10^8\msol$. Particles are colour-coded according to the density of bound particles at their individual locations ($\delta$). Comparison between collisionless (top panel) and collisional models (bottom panel) indicates that the differences introduced by the gravitational attraction of $m_\star$ are mostly confined within the thermal critical radius, $r_\epsilon$, which is marked with a solid-black circle for reference. In particular, we find that collisional haloes exhibit much higher densities within the sphere of influence of the point-mass, $r\lesssim r_\epsilon$, than the collisionless counterparts.

    This can be better seen in the middle-left panels, which show the density enhancement of field particles with negative binding energies, $\delta(r)\equiv n_h(r)/n$, in collisionless experiments. For ease of reference, dotted-black lines mark the density of the local galactic background ($\delta=1$).
    The first noteworthy result is that by plotting distances in units of the thermal critical radius, $r_\epsilon$, the profiles associated with different point-masses collapse into a single curve, thus revealing a self-similar behaviour of tidally-trapped haloes.
    Comparison with the theoretical expectation (black-dashed lines) shows that Equation~(\ref{eq:delta}) is accurate at large distances $r\gtrsim r_\epsilon$, but systematically overestimates the density of bound particles within the thermal critical radius, $r\lesssim r_\epsilon$. In particular, in their central regions halo profiles progressively depart from the power-law $\delta\sim r^{-3/2}$ and become ``cored''. The presence of a central core in the trapped halo is indeed expected from Equation~(\ref{eq:Nss_reps})
\begin{align}\label{eq:delta_reps}
  \delta(r)=\frac{1}{4\pi\,n\,r^2}\frac{\d N_{\rm ss}}{\d r}=
  \begin{cases}
    1 & ,r\lesssim r_\epsilon\\
  4/(3\sqrt{\pi})(Gm_\star)^{3/2}\sigma^{-3} e^{-V_\star^2/(2\sigma^2)}r^{-3/2} & ,  r\gtrsim r_\epsilon.
  \end{cases}
\end{align}

In contrast, bottom middle-left panel indicates that collisional models do not follow a constant-density core at small distances. Instead, the power-law `spike' $\delta\sim r^{-3/2}$ arises below the thermal critical radius, albeit with a lower normalization than outside the critical radius. As shown in the left panels of Fig.~\ref{fig:numt}, this can be traced back to the enhanced number of field particles crossing the volume $V$ in collisional models.

Top middle-right panels of Fig.~\ref{fig:prof_halo} show the one-dimensional velocity dispersion profiles of trapped particles in a steady-state. Although not shown here, we find that the trapped halo has an isotropic velocity distribution at all radii. As in previous plots, measuring distances in units of the thermal critical radius ($r_\epsilon$) removes the point-mass dependence from the profiles and yields a single, self-similar function, which follows very closely a Keplerian profile~(\ref{eq:vdisp}), $\sigma(r)\sim r^{-1/2}$ (black-dashed lines) at large distances $r \gtrsim r_\epsilon$. Again, the main differences between collisionless and collisional models are restricted to distances below the thermal critical radius. Namely, in collisionless experiments the velocity dispersion profile departs from a Keplerian power-law, converging instead towards the local velocity dispersion of the host galaxy, i.e. $\sigma_h\approx \sigma$. In contrast, collisional models do not show a convergent velocity dispersion at small distances, but rather a continuation of the Keplerian profile towards arbitrarily small distances from the point-mass.

We can now combine the density and velocity dispersion profiles measured above in order to test whether the mean phase-space density of halo particles ($Q_h=n\,\delta/\sigma_h^3$) varies with radius. Recall that the statistical theory outlined in \S\ref{sec:stats} predicts a constant phase-space density across small volume elements with a size $r\ll d$. The halo-to-field phase-space density ratio $Q_h/Q_g$ given by Equation~(\ref{eq:Qacc}) solely depends on the speed at which the point-mass travels across the Galaxy relative to the local velocity dispersion. In our experiments, $V_\star/\sigma \simeq 0.88$, which yields $Q_h/Q_g \simeq 2$. In contrast, right panel of Fig.~\ref{fig:prof_halo} that the phase-space density of bound halo particles varies with radius, and only becomes constant at large and small distances from the point-mass.
The value predicted by Equation~(\ref{eq:Qacc}) (marked with a horizontal black-dashed line) matches the empirical results at the thermal critical radius, $r\approx r_{\epsilon}$. 
Below this distance scale, $Q_h$ approaches a constant value that depends on whether or not the point-mass self-gravity is incorporated into the equations of motion. In collisionless models, the halo phase-space density of the halo converges towards the field value, $Q_h/Q_g \approx  1$ at $r\lesssim r_\epsilon$, whereas in the collisional counterparts the phase-space density ratio falls below unity, $Q_h/Q_g \lesssim 1$, which reflects the shorter dynamical lifetime of trapped particles in collisional simulations (see right panels of Fig.~\ref{fig:time}).

  It is also interesting to highlight the effect of the point-mass force on the spatial and kinematic distribution of halo particles. Namely, while collisionless simulations exhibit constant density profiles and flat velocity dispersion profiles below the thermal critical radius, collisional models follow power-law profiles $\delta\propto r^{-3/2}$ and $\sigma_h\propto r^{-1/2}$ in regions dominated by the point-mass attraction. Remarkably, both collisionless and collisional models exhibit a constant phase-space density below the thermal critical radius, $Q_h\approx {\rm const}'$ at $r\ll r_\epsilon$.


  \begin{figure}
\begin{center}
\includegraphics[width=160mm]{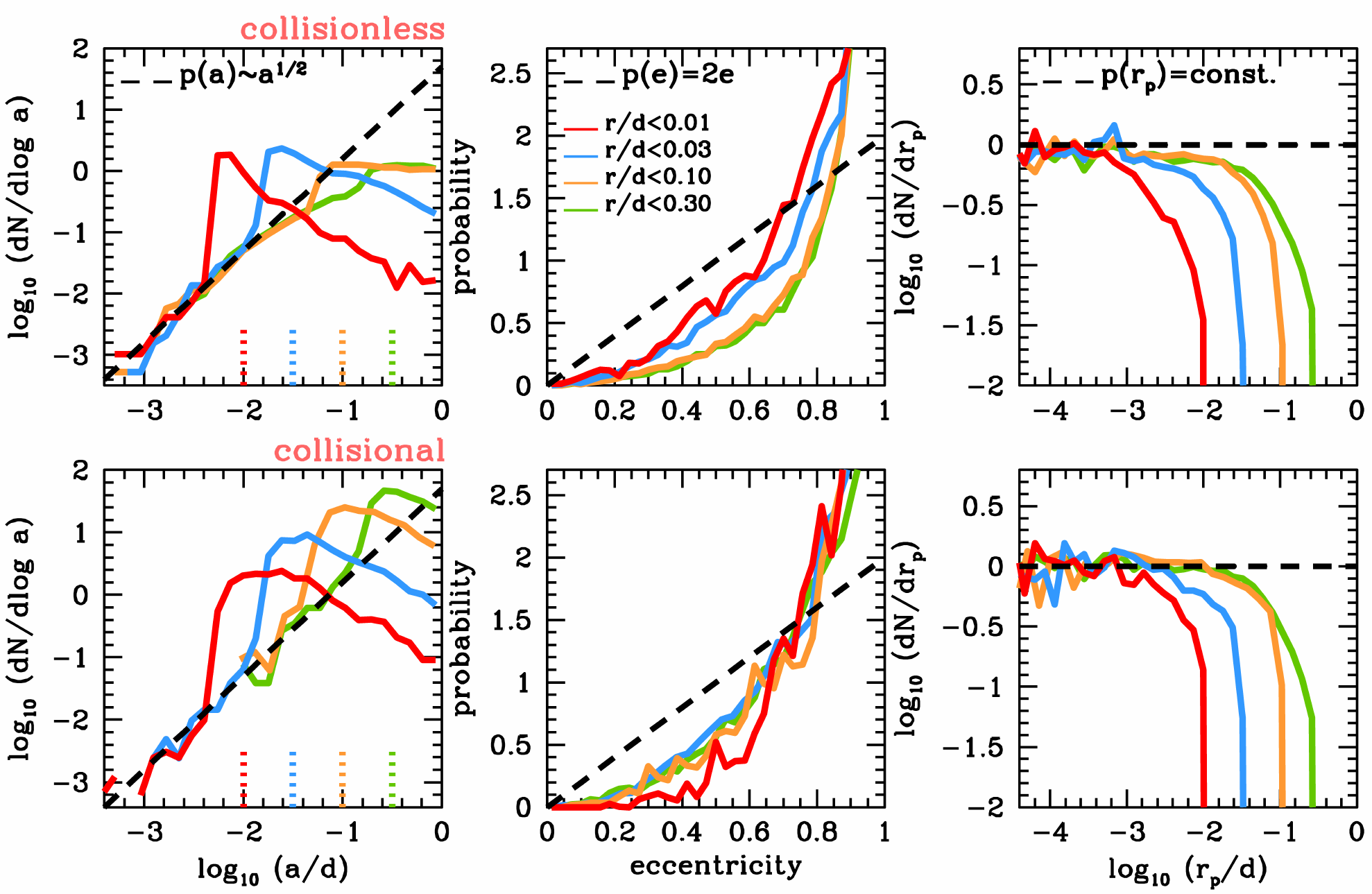}
\end{center}
\caption{{\it Left panels:} Semimajor axis distribution of particles trapped by a point-mass, $m_\star/M_g=1.3\times 10^{-4}$, within a volume $V=4\pi r^3/3$. For reference, volume sizes ($r$) are marked with vertical dotted lines. The theoretical relation~(\ref{eq:pE}) is shown with a blacked-dashed line. Note the sharp drop in the probability to find particles with a semi-major axis $a\gtrsim r/2$, which correspond to orbits with apocentres larger than the volume size, $r_a\ge 2a$.  {\it Middle panels:} Eccentricity distribution of the models shown in the left panel. The `thermal' function~(\ref{eq:pecc}) is shown with a blacked-dashed line. Regardless of volume size, all models exhibit ``super-thermal'' eccentricity distributions due to the presence of particles with large apocentres $r_a \gtrsim r$ penetrating the volume $V$ on highly-eccentric orbits. {\it Right panels:} Distribution of orbital pericentres. By construction, no particle found within the volume $V$ has orbital pericentres larger than the radius of the sphere, which translates into a truncation in the distribution $p(r_p)$ at $r_p=r$. Notice that at close distances from the point-mass, $r_p\ll r$, the probability to detect an object within a pericentre interval $r_p,r_p+\d r_p$ becomes independent of $r_p$, as expected from Equation~(\ref{eq:prperi}). }
\label{fig:dist_ae}
  \end{figure}
 
  \subsubsection{Orbits}\label{sec:orbits}
  The final task of this Section is to inspect the orbits of particles trapped in a Keplerian potential $\Phi(r)=-Gm_\star/r$. For simplicity, we compute the orbital elements of individual particles using the relative position and velocity measured at the time when the binding energy flips from a positive to a negative value, and work under the assumption that the integrals of motion are conserved quantities.
  With this simplification in mind, Fig.~\ref{fig:dist_ae} plots the semi-major axis (left panel) and eccentricity (right panel) distributions of particles accreted by a point-mass $m_\star=3\times 10^{8}\msol$ within a volume element with different sizes (marked with vertical dotted lines for reference). For a better understanding of the effect of the point-mass self-gravity, we show collissionless and collisional modes in the upper and bottom panels, respectively.
  
  In all our models, the semi-major axis distribution exhibits at prominent peak at $a=r/2$, which for radial orbits corresponds to apocentre distance that equals the volume size, $r_a=2a=r$. We find that the numerical curves approach the scale-free behaviour $p(a)\sim a^{1/2}$ predicted by~(\ref{eq:pa}) at semi-major axes that are comparable or smaller than half of the volume size, i.e. $a\ll r/2$. The probability to find particles with larger semi-major axes drops significantly, but it is far from zero, which means that some particles found at this instant of time within the volume $V$ move on orbits that take them in and out of the region under observation, i.e their orbital aponcentres are larger than the volume size, $r_a>r$. 
  
  These particles move on highly eccentric orbits, which leads to a ``super-thermal'' eccentricity distribution in the right panel of Fig.~\ref{fig:dist_ae}. In particular, there is a significant excess of halo particles with $e\gtrsim 0.7$ with respect to the thermal distribution~(\ref{eq:pecc}), $p(e)=2e$. As expected, the eccentricity distribution progressively thermalizes as the volume size shrinks and the population of particles with large apocentres $r_a>r$ decreases. 

  Comparison between collisionless and collisional models highlights the effect of incorporating the point-mass self-gravity into the equations of motion~(\ref{eq:eqmot}). Notice first that the fraction of particles with apocentres $r_a=2a\gtrsim r$ shows a systematic increase in collisional models with respect to the collisionless counterparts. This is in line with the results plotted in Fig.~\ref{fig:numt}, which show that the gravitational attraction of $m_\star$ boosts the number of objects entering the volume $V$ from distant regions on very eccentric orbits. As a result, the eccentricity distribution becomes {\it less} thermal as the volume size shrinks and the gravitational attraction of the point-mass becomes dominant, thus reversing the trend found in collisionless simulations. 

  Right panel of Fig.~\ref{fig:dist_ae} shows the distribution of orbital pericentres in these models, $r_p=a(1-e)$. As expected from Equation~(\ref{eq:prperi}), we find that the probability to find trapped particles in the range $r_{p},r_{p}+\d\,r_p$ becomes approximately constant for orbital pericentres much smaller than the volume size, $r_p\ll r$. This result holds independently of whether the attraction of the point-mass is included in the equations of motion, which has important implications for the detection of trapped interstellar objects in the solar system, as discussed in \S\ref{sec:discussion}.

  \subsubsection{Galactocentric distance}\label{sec:galactocentric}
  The above experiments place point-masses on circular orbits at a fixed galactocentric distance from the galactic potential, $R=0.065~R_0$, where $R_0$ is the host scale radius (see \S\ref{sec:setup}). In this Section, we vary the orbital radii of the point-mass in order to evaluate how the halo size changes across the host galaxy. To characterize the halo size, we measure the steady-state number of bound particles enclosed within tidal radius of the point-mass, which is computed as (e.g. Renaud et al. 2011; Pe\~narrubia et al. 2016)
\begin{align}\label{eq:rt_g}
r_t(R)=\bigg[\frac{G m_\star}{\gamma(R)\,\Omega^2(R)}\bigg]^{1/3}=\bigg[\frac{ m_\star}{\gamma(R)\,M_g(<R)}\bigg]^{1/3}R,
\end{align}
where $\gamma(R)=|\d \log n/\d \log R|$ is the power-law slope of the host's density profile, and $\Omega^2(R)=G M_g(<R)/R^3$ is the circular frequency about a host galaxy with a mass profile $M_g(<R)=4\pi\int_0^R\d r \,r^2 \,n(r)$. \S\ref{sec:tests} uses Dehnen spheres to model the galactic potential. These objects have logarithmic density slopes that approach $\gamma(R)\to \gamma$ in the limit $R\to 0$, and $\gamma(R)\to 4$ in the limit $R\to \infty$. Note that in a Keplerian potential the power-law index and the enclosed mass are radially-independent quantities, $\gamma=3$ and $\Omega^2 = G M_g/R^3$, such that Equation~(\ref{eq:rt_g}) reduces to the well-known Jacobi radius $r_t = [m_\star/(3M_g)]^{1/3}R$.

As discussed in Section~\ref{sec:stats}, the statistical theory is only accurate insofar as the local approximation can be applied, which demands the tidal radius to be much smaller than the scale length associated with the radial density profile of the host, i.e. $r_t\ll d$. It is straightforward to show that this condition fails within small distances from the galaxy centre, $R\le R_d$. For a Dehnen (1993) galaxy model, the condition $r_t(R_d)=d(R_d)$ can be solved analytically using~(\ref{eq:d_dehn}) and~(\ref{eq:rt_g}), which yields
\begin{align}\label{eq:R_d}
R_d=\bigg(\gamma^2\frac{m_\star}{M_g}\bigg)^{1/(3-\gamma)}R_0,
\end{align}
where $M_g=4\pi\,\rho_0\,R_0^3/(3-\gamma)$ is the total halo mass. For low-mass point-masses, $m_\star\ll M_g$, Equation~(\ref{eq:R_d}) returns $R_d\ll R_0$. It is thus clear that the local approximation fails at small distances from the centre of the host galaxy\footnote{Notice the special case of host galaxies with a constant density profile $\gamma=0$. In these systems the problem is not well defined due to centrally-divergent behaviour of the tidal radius, $r_t\to \infty$ at $r\to 0$, see Renaud et al. (2011) for a detailed discussion.}.
  
Fig.~\ref{fig:Nrt} plots the number of energetically-bound particles in steady-state for point-masses moving on circular orbits at different galactocentric radii. For reference, the numbers predicted by Equation~(\ref{eq:Nss_asym}) are shown with black-dashed lines. The number of trapped objects declines with galactocentric radius, which simply reflects the decreasing phase-space density of galactic particles as the distance increases. 
As expected, we find a mismatch between the theoretical curves and the numerical values at small galactocentric radii. In particular, Equation~(\ref{eq:Nss_asym}) overpredicts the number of trapped halo particles at $R\lesssim 3 R_d$, where $R_d$ is given by Equation~(\ref{eq:R_d}) and marked with vertical arrows for reference. Note that since $R_d\to 0$ as $m_\star\to 0$, the range of galactocentric distances where the theory is accurate increases as the perturber mass decreases.

  \begin{figure}
\begin{center}
\includegraphics[width=84mm]{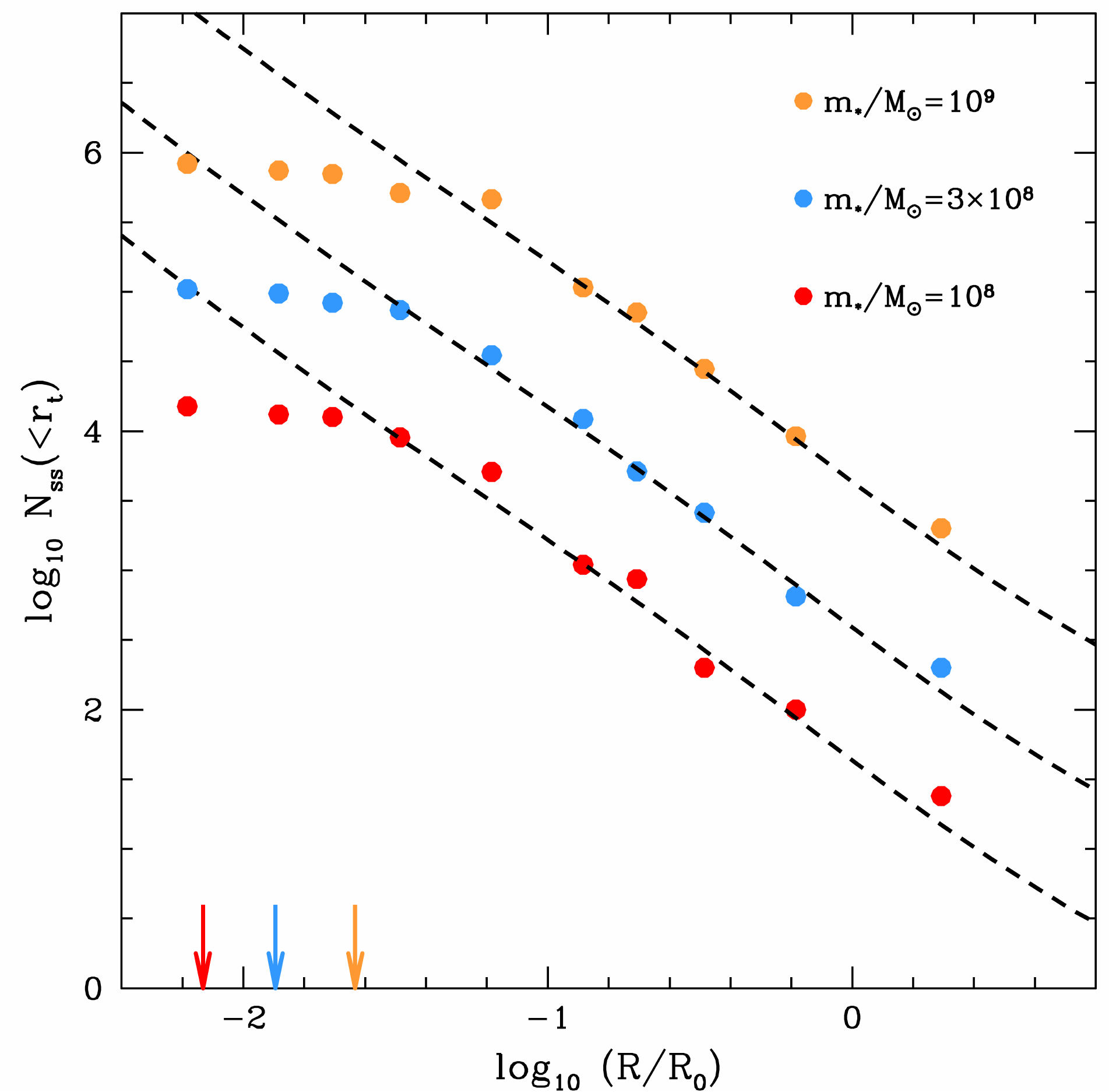}
\end{center}
\caption{Steady-state number of energetically-bound particles located within the tidal radius of point-masses moving on circular orbits at different galactocentric distances ($R$). Here, $R$ is measured in units of the host scale-length $R_0$ (see \S\ref{sec:setup}). Black-dashed lines show the values predicted by Equation~(\ref{eq:Nss_asym}). Note that the theoretical curves over-predict the number of trapped particles at distances $R\lesssim 3 R_d$, where $R_d$ corresponds to the galactocentric distance at which the point-mass tidal radius $r_t$ is equal to the local scale-length $d$ (values derived from Equation~(\ref{eq:R_d}) are marked with arrows).}
\label{fig:Nrt}
  \end{figure}
  
\section{Discussion: Trapped matter in the solar system}\label{sec:discussion}
According to the statistical theory presented above, the solar system must be embedded in an extended `halo'' of interstellar objects temporarily trapped on energetically-bound orbits around the Sun.
This Section provides a rough estimate of the steady-state number of interstellar objects and Dark Matter (DM) particles in the halo, and how they are distributed across the solar system. It is worth noting that these estimates contain a significant number of uncertainties and must be therefore viewed as preliminary steps in an ongoing effort to solve a difficult problem. For example, the statistical analysis neglects the response of trapped halo particles to interactions with planets as well as Galactic substructures in the vicinity of the Sun, which could potentially affect the numerical values derived below (see Section~\ref{sec:planets}).

For the estimates below, the Sun is placed at ${\bb R}_\odot =(-8.17,0.0,0.02)$ kpc (Gravity Collaboration et al. 2019). At this location, the circular velocity is $V_c(R_\odot)=237\kms$ (McMillan et al. 2017) and the Local Standard of Rest (LSR) is ${\bb V}_{\rm LSR}=(11.1,12.2,7.3)\kms$ (Sch\"oenrich et al. 2010). The Galactic potential is modelled as a  disc, bulge and dark matter halo with the following parameters. The disc is a Miyamoto-Nagai (1975) model with a mass $M_d=6.6\times 10^{10}M_\odot$, and radial and vertical scale lengths $a=8$ kpc and $b=0.3$ kpc. The MW bulge follows a Hernquist (1990) profile with a
mass $M_b=2.3\times 10^{10}M_\odot$ and a scale radius $c=1.2$ kpc.
The MW dark matter halo is modelled as a Navarro, Frenk \& White (1997) profile with a virial mass $M_{\rm vir}=1.2 \times 10^{12}M_\odot$, virial radius $r_{\rm vir}=222$ kpc and scale-radius $r_s=14.1\kpc$ (Bland-Hawthorn \& Gerhard 2016). In this model, the velocity dispersion of the DM halo at the solar radius is $\sigma_{\rm DM}\simeq 135\kms$.
For the stellar velocity dispersion in the Solar vicinity we adopt the values measured by Anguiano et al. (2020) using Gaia DR2 and APOGEE data, who find $\sigma_{\rm thin}\simeq 28\kms$, $\sigma_{\rm thick}\simeq 54\kms$ and $\sigma_{\rm halo}\simeq 120\kms$, for the thin and thick discs and the stellar halo, respectively. For simplicity, we will assume that interstellar particles move on isotropic trajectories. 

  \begin{figure}
\begin{center}
\includegraphics[width=160mm]{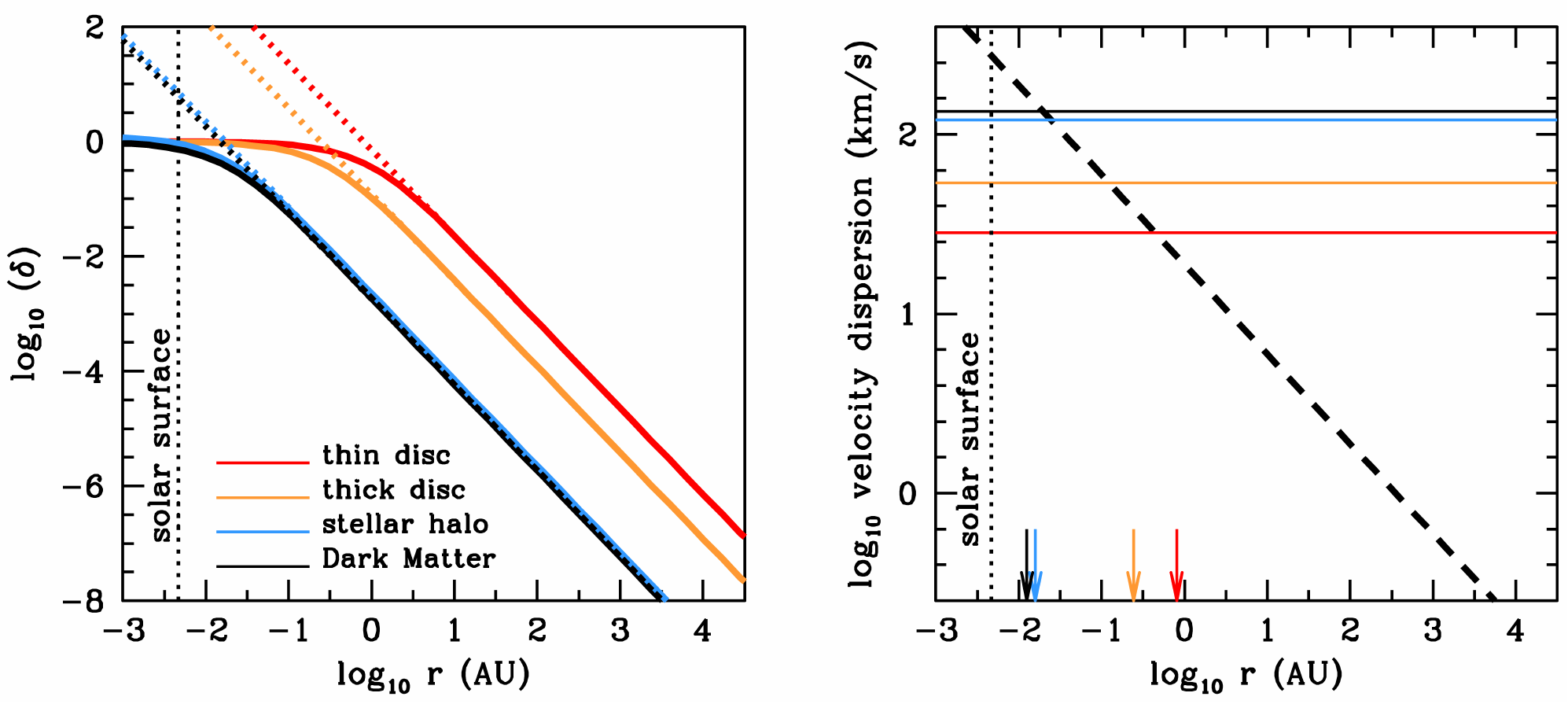}
\end{center}
\caption{{\it Left panel:} Density enhancement, $\delta(r)\equiv n_h(r)/n$, as a function of distance from the Sun for baryonic ISOs and DM particles. Dotted lines correspond to the analytical formula~(\ref{eq:delta}), whereas solid lines show the empirical curves found in collisionless models, Equation~(\ref{eq:delta_reps}). The Solar radius is marked with a vertical-dotted line for reference. {\it Right panel:} Velocity dispersion profile of ISOs and DM particles bound to the Solar potential (black dashed lines). For reference, coloured-horizontal lines show the velocity dispersion of each Milky Way component considered in the analysis. Arrows mark the radii where $\delta=1$, Equation~(\ref{eq:r_eps}). }
\label{fig:est_MW}
  \end{figure}

\subsection{Halo composition} \label{sec:ss}
As the Sun moves through the Milky Way, it traps light bodies that populate the interstellar background (comets, rocks, asteroids, DM particles, etc). The frequency of the entrapment events strongly depends on the kinematics of these objects in the vicinity of the solar system.

Left panel of Fig.~\ref{fig:est_MW} plots the density enhancement of halo objects ($\delta\equiv n_h/n$) as a function of distanced from the Sun.
Lines are colour-coded according to whether ISOs are tidally injected from the thin and thick discs, and the stellar and Dark Matter (DM) haloes. 
Dotted and solid lines correspond to Equation~(\ref{eq:delta}) and the empirically-corrected expression~(\ref{eq:delta_reps}), respectively. Recall that these estimates adopt $\alpha=1$, which is a conservative choice at distances $r\gtrsim r_\epsilon$ (see numerical experiments plotted in the right panel of Fig.~\ref{fig:time}).
As expected, tidal trapping is most efficient for bodies that belong to the thin disc (red lines), which co-rotate with the Sun and have a low velocity dispersion.
Interestingly, the radius where the population of trapped ISOs dominates over the number of (unbound) Milky Way interlopers roughly coincides with the radius of the Earth's orbit, $r_\epsilon^{\rm ISO}\approx 0.8\AU$. Therefore, {\it the majority of ISOs reaching the inner-most regions of the solar system will be energetically bound to the Sun, and by definition members of the trapped halo}.

In addition, the halo of trapped matter also contains a dark matter component. The DM density enhancement is relatively low on account on the large velocity dispersion of DM particles ($\sigma_{\rm DM}\simeq 135\kms$) and the fact that the Sun is speeding through the Galactic halo at $\sim 245\kms$. According to the estimates plotted in Fig.~\ref{fig:est_MW}, the density of bound DM particles exceeds the local DM background in the proximity of the Solar surface, $r_\epsilon^{\rm DM}\approx 0.015\,{\rm AU}\simeq 3\,r_\odot$, where $r_\odot$ is the nominal Solar radius, suggesting that the vast majority of DM particles cross the solar system on unbound trajectories.

Right panel of Fig.~\ref{fig:est_MW} plots the velocity dispersion of particles trapped in the halo (black-dashed line) as a function of distance from the Sun, Equation~(\ref{eq:vdisp}). In our statistical model, all bound particles follow the same dispersion profile regardless of their Galactic origin. In this sense, the process of tidal trapping erases any dynamic `memory' of their past orbits in the Galactic potential. Comparison with the velocity dispersion of the Milky Way components (plotted with horizontal lines for reference) shows that the halo  of trapped objects is much `colder' than the Galactic environment at $r\gtrsim 1\AU$. At smaller distances, the halo velocity dispersion raises towards the Sun as $\sigma_h\sim r^{-1/2}$, exceeding the background velocity dispersion at distances comparable to the thermal critical radius, $r\lesssim r_\epsilon$, where $r_\epsilon$ is given by Equation~(\ref{eq:r_eps}) (marked with vertical arrows for ease of comparison). Recall that in collisionless approximation, the radius at which the density enhancement reaches unity coincides with the thermal critical radius, i.e. $\delta(r_\epsilon)=1$ (see \S\ref{sec:profiles}).

  \subsection{Interstellar Objects} \label{sec:ss}
In our theory, the number of tidally-trapped ISOs is determined by the local background density and the velocity dispersion of these objects in the solar vicinity. 
The discoveries of the irregular body 'Oumuamua (Meech et al. 2017) and the comet Borisov (Jewitt \& Luu 2019), both travelling on a unbound orbits, suggests that the local number density of ISOs is of the order of $n^{\rm ISO}\sim 2\times 10^{15}\pc^{-3}$ (Do et al. 2018; Jewitt et al. 2020).
The local velocity distribution of ISOs is still unknown. Here, we use the estimates plotted in Fig.~\ref{fig:est_MW}, which suggest that the contribution from the thick disc and stellar halo to the population of trapped ISOs can be neglected at leading order, such that $\sigma^{\rm ISO}\approx\sigma_{\rm thin}$.

Left panel of Fig.~\ref{fig:n_m_v} shows the expected number of ISOs as a function of distance to the Sun. For reference, we plot the cumulative number of visitors expected from a constant-density background, $N^{\rm ISO}(r)=4\pi n^{\rm ISO}r^3/3$, is plotted with a blue line, while the steady-state number of energetically-bound ISOs (orange line) is estimated from Equation~(\ref{eq:Nss_reps}) as
 \begin{align}\label{eq:Nss_est}
  N_{\rm ss}^{\rm ISO}(r)\approx 4\times 10^6 \,\bigg(\frac{m_\star}{M_\odot}\bigg)^{3/2}\bigg(\frac{n^{\rm ISO}}{2\times 10^{15}\pc^{-3}}\bigg)\bigg(\frac{\sigma^{\rm ISO}}{28\kms}\bigg)^{-3}\bigg(\frac{r}{2\times 10^4\AU}\bigg)^{3/2}
 \end{align}
 at $r\gg r_\epsilon$. We stress that the estimate~(\ref{eq:Nss_est}) becomes uncertain beyond the region populated by Oort comets, $r\gtrsim 2\times 10^4\AU$, which is poorly constrained observationally and may be subject to Galactic perturbations (see \S\ref{sec:planets}).

 According to~(\ref{eq:Nss_est}), there are of the order of $N_{\rm ss}^{\rm ISO}(30\AU)\sim 226$ `Oumuamua-like objects trapped in the solar system at distances comparable to the orbit of Neptune. 
Interestingly, Namouni \& Morais (2020) found several Centaurs and a few Trans Neptunian objects that may have been captured from the interstellar background, which would go in line with our theoretical predictions (but see Morbidelli et al. 2020).

The population of bound ISOs is dwarfed by the number of Milky Way interlopers currently passing through the solar system, which exceeds $N^{\rm ISO}\sim 10^{13}$ at $r\sim 2\times 10^4\AU$. In contrast, the Oort Cloud is thought to host some $\sim 10^{11}$ energetically-bound comets (Oort 1950; Heisler 1990; Brasser \& Morbidelli 2013; Feng \& Bailer-Jones 2014; Rickman et al. 2017).
In agreement with Siraj \& Loeb (2021), our estimates suggest that unbound Milky Way interlopers may outnumber bound Oort cloud members by roughly $\sim 2$ orders of magnitude.
Note, however, that the size and spatial distribution of the Oort Cloud is a long-standing matter of debate (for reviews see Dones et al. 2004; Rickman 2014; Dones et al. 2015; Jewitt \& Seligman 2022). Direct observations in this region are difficult due to the vast distances involved. Current estimates can only (indirectly) inferred by comparing the predictions from planet formation models against the (infrequent) visits of long-period comets to the inner solar system, or from the modelling of the cratered surfaces of the Moon and terrestrial planets.
Existing estimates on the number of objects in the Oort cloud are model dependent and still uncertain because the flux of long-period comets that reach the inner solar system with pericentres $r_p < 4 \AU$ is only 2--3 per year, so an error of only 1 comet per year can significantly affect the existing constraints (e.g. Moro-Mart\'in 2019).

Crucially, existing estimates on the number of objects in the Oort Cloud do not account for the presence of tidally-trapped ISOs in the solar system, which are expected to contribute to the population of minor bodies with large apocentres and negative binding energies. For example, Figs.~\ref{fig:est_MW} and~\ref{fig:n_m_v} indicate that most ISOs detected at $r\lesssim 1\AU$ will be bound to the Sun's potential and may not be easily distinguishable from genuine members of the Oort Cloud. This issue has been recently discussed by Dehnen \& Hands (2022), who also point out that captured ISOs follow orbits akin to those of known long-period comets, and that some of these comets could be of extra-solar origin. 
These results open up interesting questions regarding the composition of the outer solar system that may be worth investigating in a separate contribution.

  \begin{figure}
\begin{center}
\includegraphics[width=168mm]{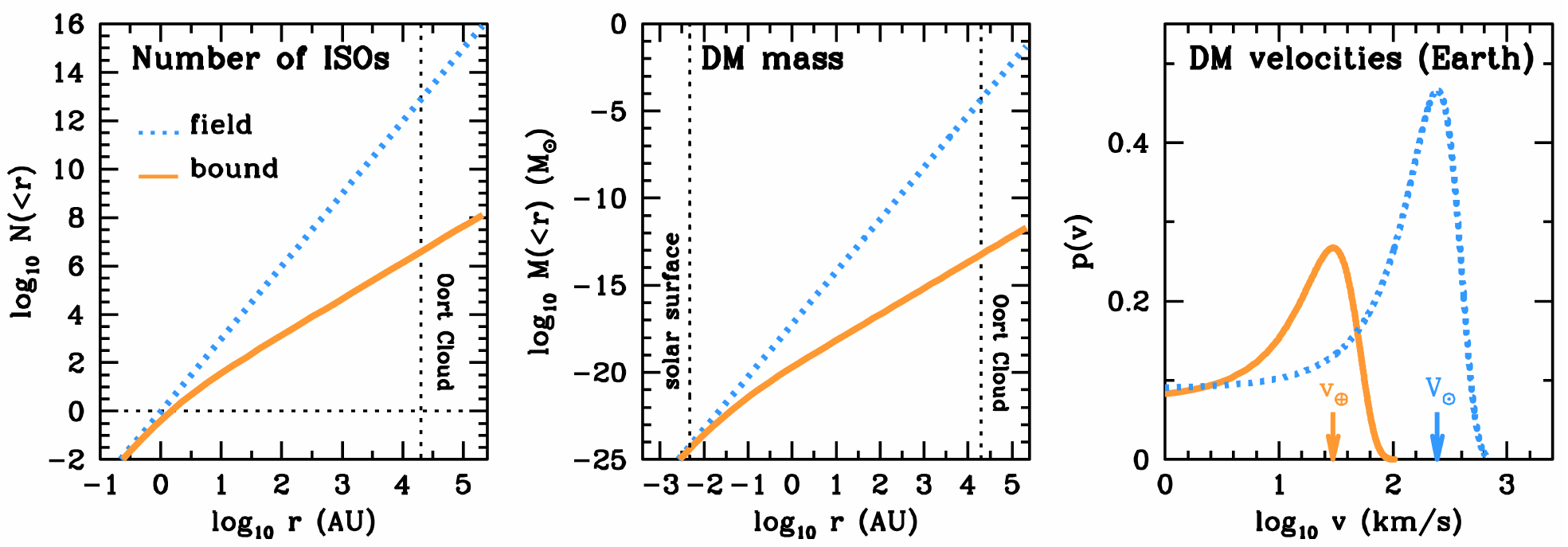}
\end{center}
\caption{{\it Left panel:} Number of (baryonic) Interstellar Objects (ISOs) as a function of distance from the Sun estimated from the number density and velocity dispersion of ISOs in the Solar vicinity (see text). 
  Orange line plots the number of bound objects derived from Equation~(\ref{eq:Nss_reps}). For reference, dotted-blue line shows the number of interlopers that one would expect from the (unperturbed) field, $N=4\pi n r^3/3$. Horizontal and vertical dotted lines mark $N=1$ and the canonical distance to the Oort Cloud, $2\times 10^4\,{\rm AU}$, respectively. Notice that most ISOs crossing the radius $r\lesssim 1\AU$ are expected to be energetically bound to the sun. {\it Middle panel:} Cumulative Dark Matter mass in the solar system as a function of distance from the Sun. {\it Right panel:} Maxwellian velocity distribution of Dark Matter particles expected at the Earth location. The presence of a bound DM halo surrounding the solar system manifests as a narrow velocity peak centred at the Earth-Sun relative velocity, $v_\oplus\simeq 29.7\kms$, which is marked with an orange arrow for reference (see text). In contrast, unbound Milky Way dark matter particles crossing the solar system have a much wider velocity distribution that peaks at the relative velocity between the Sun and the Milky Way centre, $V_\odot\simeq 245\kms$.
}
\label{fig:n_m_v}
  \end{figure}

\subsection{Dark Matter}\label{sec:DM}
The Sun can also trap DM particles from the Milky Way DM halo.
For our estimates, we adopt a background DM density of $\rho_{\rm DM}=0.012 \,M_\odot/\pc^3$ (e.g. Read 2014, and references therein), and a velocity  dispersion $\sigma_{\rm DM}=135\kms$.
Middle panel of Fig.~\ref{fig:n_m_v} plots the DM mass profile as a function of heliocentric distance, with orange/blue lines denoting bound/unbound particles, respectively. This plot highlights a number of interesting results. First, notice first that close to the Solar surface, $r\lesssim r_\epsilon^{\rm DM}\approx 0.015\AU$, all DM particles are energetically-bound to the Sun, and are therefore members of the bound halo.  This result may be relevant in the context of capture of DM by the Sun, which is typically estimated by adopting a velocity distribution that describes the (unbound) Milky Way DM component at the solar location (for a review, see Nu{\~n}ez-Casti{\~n}eyra et al. 2019), thus ignoring the effect of the Sun's gravity on the trajectories of DM particles as they approach the solar system.

At larger radii, $r\gg r_\epsilon^{\rm DM}$, the halo mass in bound DM particles can be derived from~(\ref{eq:Nss_reps}) as
 \begin{align}\label{eq:Mss_est}
  M_{\rm h}^{\rm DM}(r)\approx 6\times 10^{-14}\,M_\odot \,\bigg(\frac{m_\star}{M_\odot}\bigg)^{3/2}\bigg(\frac{\rho_{\rm DM}}{0.012\, M_\odot\pc^{-3}}\bigg)\bigg(\frac{\sigma_{\rm DM}}{135\kms}\bigg)^{-3}\bigg(\frac{r}{2\times 10^4\AU}\bigg)^{3/2},
 \end{align}
 which indicates that the amount of DM trapped in the solar system is small, $M_{\rm h}^{\rm DM}(0.1\pc)\sim 10^{-13}M_\odot$. This value sits well below current upper limits derived from the motion of planets, which constrain the halo mass within Saturn's orbit below $<1.7\times 10^{-10}M_\odot$ (Pitjev \& Pitjeva 2013).
 The presence of trapped DM particles induces an extra gravitational attraction of $\sim 2\times 10^{-23}{\rm m}/{\rm s}^2$ at $r\sim 70\AU$, which cannot explain the so-called Pioneer anomaly, a uniform acceleration of $\sim 8\times 10^{-10}{\rm m}/{\rm s}^2$ towards the Sun detected by spacecrafts Pioneer 10 and 11 at a distance between $20\AU$ and $70\AU$ (Turyshev \& Toth 2010 and references therein).

 On Earth, the density of bound DM particles is subdominant with respect the Galactic background. According to Equation~(\ref{eq:delta_reps}) the density enhancement induced by the bound DM halo is $\delta^{\rm DM}(r_\oplus)\approx 1.8\times 10^{-3}$ at $r=r_\oplus=1\AU$. However, despite their relatively small contribution to the DM density on Earth, trapped DM particles have a relatively high phase-space density. This property is key for direct detection attempts using elastic atomic-recoil experiments (Lewin \& Smith 1996), as well as to estimate capture rates of DM particles by the Earth, which are expected to self-annihilate in the core and produce neutrinos that can be detected (e.g. Gould 1987, 1988).

This issue is illustrated in the right panel of Fig.~\ref{fig:n_m_v}, which shows the relative velocity distribution of DM particles expected at $r=1\AU$, which under the Maxwellian approximation~(\ref{eq:pv}) can be written as
  \begin{align}\label{eq:pv_est}
    p({\bb v|r_\oplus})=p_{\rm h}({\bb v|r_\oplus})+p_{\rm MW}({\bb v|r_\oplus})=\frac{\delta(r_\oplus)}{[2\pi \sigma_h^2(r_\oplus)]^{3/2}}\exp\bigg[-\frac{( v- v_\oplus)^2}{2\sigma_h^2(r_\oplus)}\bigg] + \frac{1}{(2\pi \sigma_{\rm DM}^2)^{3/2}}\exp\bigg[-\frac{(v-V_\odot)^2}{2\sigma_{\rm DM}^2}\bigg],
  \end{align}
  where the velocity vectors have been aligned with the Solar motion for ease of comparison.
In Equation~(\ref{eq:pv_est}), the left-hand term is due to the presence of tidally-trapped DM particles on Earth, which manifest in a cold velocity peak centred at the Earth-Sun velocity, $v_\oplus=29.7\kms$ (marked with an orange arrow for reference) with a dispersion $\sigma_h(r_\oplus)=(2/5)^{1/2}v_\oplus\approx 18.8\kms$ given by~(\ref{eq:vdisp}). The right-hand term accounts for Milky Way DM particles crossing the solar system on unbound trajectories, and exhibits a much wider velocity distribution centred at the Galactocentric velocity, $V_\odot=245\kms$ (blue arrow) with a dispersion $\sigma_{\rm DM}=135\kms$.
The relative amplitude between the two velocity peaks is set by the mean phase-space density of the halo relative to the background
$$\frac{p_{\rm h}({\bb v_\oplus|r_\oplus})}{p_{\rm MW}({\bb V_\odot|r_\oplus})}=\delta(r_\oplus)\frac{\sigma^3_{\rm DM}(R_\odot)}{\sigma^3_{\rm h}(r_\oplus)}=\frac{Q_h(r_\oplus)}{Q_g}.$$
  From Equation~(\ref{eq:Qacc}) we find $Q_h/Q_g\simeq 0.6$ at the Earth location.  Furthermore, since the Earth moves on a nearly circular orbit around the Sun, experiments that aim to detect bound DM particles should expect no annual modulation.
  Crucially, current exclusion limits on the DM particle mass \& cross section derived from atomic-recoil experiments \& neutrino telescopes neglect the cold velocity peak generated by tidally-trapped particles shown in Fig.~\ref{fig:n_m_v}, and may therefore be more sensitive to higher particle masses than previously thought.

\subsection{Collisional effects}\label{sec:planets}
For simplicity, our analysis ignores the collisional element of the problem.
E.g., the above estimates do not account for collisions between infalling stellar particles and the planets (or any other massive object) orbiting around the Sun. It also ignores gravitational perturbations from Galactic substructures in the vicinity of the solar system, either visible (e.g. stars \& gas clouds) or invisible (black holes, free-floating planets, dark matter subhaloes, etc).

Galactic substructures generate random fluctuations of the local tidal field that inject kinetic energy onto bound orbits (e.g. Chandrasekhar 1941; Pe\~narrubia 2018). Over time, the cumulative effect of tidal fluctuations causes tidal heating and a progressive erosion of the Oort cloud (Chandrasekhar 1941; Hut \& Tremaine 1985; Pe\~narrubia 2019; Torres et al. 2019; Portegies Zwart 2021; Portegies Zwart et al. 2021). For example, Monte-Carlo experiments carried by P21\footnote{Note that P21 experiments follow the evolution of wide binaries in a clumpy potential. However, the results are applicable here on account that the relative motion between binary stars is governed by the same differential equations that determine the orbits of tracer particles orbiting around a Keplerian potential. }  show that tidal evaporation truncates the semi-major axis distribution of halo particles, and that the truncation is smaller than the tidal radius of the Keplerian potential (c.f. Fig.~8 in P21).
At present, it is unclear whether the solar system extends beyond the Oort Cloud, which calls for caution when extrapolating the results of \S\ref{sec:ss} and \S\ref{sec:DM} at radii $r\gtrsim 2\times 10^4\AU$.

The role of planets in shaping the population of tidally-trapped objects is more difficult to forecast, as their gravitational attraction induces competing effects.
On the one hand, interactions with planets may inject energy onto the orbits of minor bodies orbiting around the Sun, thereby shortening their dynamical lifetimes (e.g. Yabushita 1980; Napier et al. 2021b). Yet, in the case of tidally-trapped objects the number of hard encounters is limited by the short time span that these objects spend in the solar system, which is comparable to the duration of a fly-by encounter, $t_{\rm surv}\sim T$ (see Fig.~\ref{fig:time}).
This picture changes for semi-stable captures with long survival times, $t_{\rm surv}\gg T$, which may experience a significant amount of heating if their orbits bring them into the regions of the solar system populated by planets.

On the other hand, 3-body collisions with planets provides an independent mechanism for capturing interstellar visitors onto bound orbits, thus contributing to the bound halo population.
For example, Hands \& Dehnen (2020) simulate $\sim 4\times 10^8$ trajectories of interstellar objects that approach a solar system in isolation (i.e. external forces are ignored), and finds a steady-state population of up to $\sim 10^4$--$10^5$ `Oumuamua-style objects, most of them located at $r\lesssim 100\AU$ from the Sun (e.g. Fig.~5 of Dehnen et al. 2021).

The complex interplay between the above dynamical processes calls for follow-up experiments that include collisional terms in the equations of motion~(\ref{eq:eqmot}).

\section{Summary }
This paper shows that Galactic tidal forces can inject ISOs and DM particles onto temporarily-bound orbits around the Sun.
The mechanics of tidal injection are akin to slingshot manoeuvres, where the motion of the infalling object in the Galactic potential combines with the gravity of the Sun to decelerate the relative speed an infalling body to a degree where the binding energy becomes negative. 
The same mechanism responsible for capturing interstellar particles also leads to their later unbinding.
As a result, tidally-trapped objects have a limited life span in the solar system before being lost back to the Galactic potential.

The presence of tidally-trapped objects creates an overdensity of interstellar material surrounding the point-mass, or ``halo'', which reaches a steady state as the number of objects tidally injected from the field becomes comparable to those being tidally stripped. It should be stressed that the halo does not consist of the same bodies which are permanently captured, but rather of temporarily-trapped bodies moving through the overdensity with negative binding energies.

The stochastic model presented in this paper applies Smoluchowski (1916)'s kinetic theory to derive the entrapment rate and the steady-state phase-space distribution of tidally-trapped particles in a Keplerian potential, $\Phi_\star(r)=-Gm_\star/r$. In a steady state, bound particles generate a local density enhancement $\delta(r)\sim r^{-3/2}$ (also known as a `density spike') and follow an isotropic velocity dispersion profile that scales as $\sigma_h(r)\sim r^{-1/2}$. 
Tests against live $N$-body models that allow the Milky Way to respond to the gravitational attraction of a point-mass $m_\star$ show that the analytical derivations are accurate on distance range $r\gtrsim r_\epsilon$. Here, $r_\epsilon$ is the thermal critical radius, Equation~(\ref{eq:r_eps}), which marks the distance from a point-mass at rest where the Keplerian potential roughly equals the mean kinetic energy of field particles, $|\Phi_\star(r_\epsilon)|\approx 3\sigma^2/2$. Numerical experiments show that the stochastic theory tends to over-estimate the density of trapped particles below this scale (see \S\ref{sec:tests}).

It is worth noting that although tidal trapping does not require additional mechanism(s) to remove kinetic energy from infalling interstellar material, 3-body interactions with planets, resonant capture and friction in the natal gas cloud will also contribute to the population of bound interstellar particles surrounding the solar system. Alas, these dynamical processes are strongly intertwined, which adds significant complexity when making predictions on the amount and composition of the interstellar matter currently bound to the Sun.

Our theory predicts that other massive bodies travelling through the Milky Way must also be embedded in tenuous envelopes of trapped interstellar particles.
Here we focus on the solar system, but similar arguments can be extended to individual planets (e.g. Adler 2009), as well as Black Holes moving through the Galaxy (e.g. Gondolo \& Silk 1999; Boudaud et al. 2021). This possibility will be exploring in separate contributions.

\section*{Acknowledgements}
This work has greatly benefited from the insightful comments of Malcolm Fairbairn, Henrique Araujo, Michael Petersen and Raphael Errani. The author would also like to thank the anonymous referee for very helpful suggestions.
 
\section*{Data Availability}
No data were generated for this study.

{}

\appendix
\section{Density profile implied by a constant Distribution Function}

In order to derive density and velocity dispersion profiles from a distribution function, one needs to compute the velocity element 
\begin{align}\label{eq:d3v1}
  \d^3 v= 2\pi v_t \d v_t \d v_r=2\pi \frac{L \d L}{r^2}\frac{\d E}{v_r},
\end{align}
where $v_r=\{2[E+Gm_\star/r-L^2/(2r^2) ]\}^{1/2}$ and $v_t=L/r$ are the radial and tangential components of the velocity vector, respectively. For particles moving on a Keplerian orbit the condition $v_r=0$ is met at $r_{\rm a}=a(1+e)$ (apocentre) and $r_{\rm p}=a(1-e)$ (pericentre). Hence,
\begin{align}\label{eq:vr}
  v_r^2=2\bigg[-\frac{Gm_\star}{2a}+\frac{Gm_\star}{r}-\frac{Gm_\star a(1-e^2)}{2r^2}\bigg]=\frac{Gm_\star}{a}\bigg(\frac{a}{r}\bigg)^2\bigg[\frac{r}{a}-(1-e)\bigg]\bigg[(1+e)-\frac{r}{a}\bigg].
\end{align}
Inserting~(\ref{eq:vr}) into~(\ref{eq:d3v1}) and expressing energy and angular momentum as a function of semimajor axis and eccentricity yields
\begin{align}\label{eq:d3v}
\d^3 v=\frac{\pi (Gm_\star)^{3/2}}{r} e\d(-e)\frac{\d a}{a^{3/2}}\frac{1}{\{[r/a-(1-e)][(1+e)-r/a]\}^{1/2}}.
\end{align}

The number density of halo particles is found by integrating~(\ref{eq:DF}) over~(\ref{eq:d3v}) 
\begin{align}\label{eq:rho}
  n_h(r)&=\int \d^3 v\,f \\\nonumber
  &=\pi f_0\frac{(Gm_\star)^{3/2}}{r}\int_{0}^{\infty}\frac{\d a}{a^{3/2}}\int_0^{1}\d e\frac{e}{\{[r/a-(1-e)][(1+e)-r/a]\}^{1/2}} \\ \nonumber
  &=\pi f_0\frac{(Gm_\star)^{3/2}}{r}\int_{0}^{\infty}\frac{\d a}{a^{3/2}}\bigg\{\bigg[\frac{(2a-r)r}{a^2}\bigg]^{1/2}-\bigg[\frac{(2a-r)r-a^2}{a^2}\bigg]^{1/2}\bigg\} \\ \nonumber
&=\frac{4\pi\sqrt{2}}{3}f_0\bigg(\frac{Gm_\star}{r}\bigg)^{3/2},
  \end{align}
which recovers the profile~(\ref{eq:delta}) for
\begin{align}\label{eq:f0}
  f_0=\frac{1}{2\sqrt{2\pi}}e^{-V_\star^2/(2\sigma^2)}\,Q_g,
\end{align}
with $Q_g=n/\sigma^3$ being the mean phase-space density of the interstellar background.

It is worth noting that the power-law profile~(\ref{eq:rho}) arises from particle ensembles orbiting in a Keplerian potential with constant phase-space density. This profile has been studied by other authors in several contexts (e.g. Gondolo \& Silk 1999; Oncins et al. 2022)  

\section{Wide binaries}
It is straightforward to show that the distribution function~(\ref{eq:DF}) also describes the orbits of wide binaries formed in an homogeneous sea of particles with Maxwellian velocities.  
Expressing~(\ref{eq:pa}) as a function of orbital energy via the transformation $p(E)\,\d E= p(a)\,\d a$ with $a=Gm_\star/(-2 E)$ leads to an energy distribution
\begin{align}\label{eq:pE}
  p(E)\,\d E\sim \frac{\d E}{(-E)^{5/2}}.
\end{align}  
Hence, re-writing~(\ref{eq:DF}) in terms of energy yields
\begin{align}\label{eq:DFE}
  p(E,e)\,\d E\d e \sim \frac{1}{(-E)^{5/2}}\,(2e)\,\d E\,\d e,
\end{align}  
which matches the distribution of integrals found by Jeans (1928) (see also Heggie 1975) 
\begin{align}\label{eq:pE_Jeans}
 p_{\rm pairs}(E,e)\d E\d e\sim \frac{\exp(-E/\mathcal T)}{(-E)^{5/2}}(2e)\,\d E\,\d e,
\end{align}
in the limit where stellar pairs move on orbits with small relative velocities $v<v_e\ll \sigma$, and low binding energies, $|E|\ll \mathcal T=3\sigma^2/2$, such that $\exp(-E/\mathcal{T})\approx 1$.

\section{Numerical integration}\label{sec:con}
This Section briefly explains how the coupled differential equations~(\ref{eq:eqmotsun}) and~(\ref{eq:eqmot}) are solved in this work.
The main difficulty lies in the vastly different dynamical ranges of particles bound to the point-mass and those moving the in Galactic potential. For illustration, consider the models shown in Section~\ref{sec:tests}, which resolve distances below the critical radius of the point-mass, $r_0=2Gm_\star/\sigma^2$. At this distance, the circular frequency is $\Omega_\star(r_0)=(Gm_\star/r_0^3)^{1/2}=2^{-3/2}\sigma^3/(Gm_\star)$. This number can be directly compared with the circular frequency of the point-mass around the host galaxy centre, $\Omega_g(R_\star)=[GM_g(<R_\star)/R^3_\star]^{1/2}$. Estimating the local velocity dispersion as $\sigma^2\approx V_c^2/2=GM_g(<R_\star)/(2R_\star)$ yields a frequency ratio $\Omega_g/\Omega_\star\sim 8 m_\star/M_g(<R_\star)$.
Given that the numerical models in Section~\ref{sec:tests} have mass ratios $m_\star/M_g\ll1$, it seems crucial to solve equations~(\ref{eq:eqmotsun}) and~(\ref{eq:eqmot}) taking into account the disparate dynamical range between particles moving in the host galaxy and those temporarily bound to the Keplerian potential.

To this aim, we apply a leap-frog scheme with a varying time-step. In particular, we use a fourth-order Runge-Kutta integrator (Press et al. 1995) with a time-step that is adjusted at any point of the integration according to the motion motion of each individual particle. We define two orbital frequencies, $\omega_\star =v/r$ and  $\omega_g =V_\star/R_\star$ (recall that capital/lower letters denote quantities measured with respect to the host galaxy/point-mass centre), where the phase-space quantities $\{r,v,R_\star,V_{\star}\}$ are measured at each integration time $t$. The integration time-step at the time $t$ is then set as a fraction of the smallest of the two associated periods, $$\Delta t= \epsilon\,\cdot {\rm min}\,(\omega^{-1}_\star,\omega^{-1}_g),$$ where $\epsilon\ll 1$ is a dimension-less free parameter.
Notice that by using the same time-step to solve~(\ref{eq:eqmotsun}) and~(\ref{eq:eqmot}) both trajectories are kept synchronous throughout the integration, avoiding the need to interpolate between different time-steps. However, since $\omega_\star\gg \omega_g$, this choice means that Equation~(\ref{eq:eqmotsun}) is solved with much higher accuracy than~(\ref{eq:eqmot}).
In practice, the integration parameter $\epsilon$ is set by reaching a compromise between numerical accuracy and computational cost. To find the optimal value we study the convergence of the ensemble properties of the population of tidally-trapped particles as a function of $\epsilon$.

Fig.~\ref{fig:con} shows how varying $\epsilon$ affects the steady-state distribution of trapped particles surrounding a point-mass with $m_\star=3\times 10^8\msol$. 
 For this exercise, we use the initial conditions outlined in \S\ref{sec:setup}. Each halo is rotated $N_{\rm ens}=40$ times in random directions, which yields an effective number of particles $N_{\rm eff}=N_g\times N_{\rm ens}=2\times 10^8$. The motion of each individual particle is computed from Equations~(\ref{eq:eqmotsun}) and~(\ref{eq:eqmot}) in a `collisional' setup with $F_\star\ne 0$ for a total integration time $t_f=10\,T(d)$.
 Left, middle and right panel plot the density enhancement, velocity dispersion profile and mean phase-space density of particles with $E<0$. For comparison, black-dashed lines show the theoretical profiles~(\ref{eq:delta}),~(\ref{eq:vdisp}) and~(\ref{eq:Qacc}), respectively. 

 These tests reveal how the choice of $\epsilon$ affects the spatial and kinematic distributions of trapped particles. Convergence arises for values $\epsilon \lesssim 3\times 10^{-3}$. Large values of the integration parameter $\epsilon$ --and therefore longer time-steps -- do not capture the dynamics of trapped particles. More precisely, large values of the integration parameter $\epsilon$ --and therefore longer time-steps-- lead to density profiles that over/underestimate the number of bound particles beyond/within $r_\epsilon$. Interestingly, the velocity dispersion profiles barely show any sensitivity to the choice of $\epsilon$.
 
 The impact of $\epsilon$ on our models is best appreciated in the right panel, which shows the radial variation of the halo mean phase-space density, $Q_h=\delta\,n/\sigma^3_h$, normalized by the local value of the field, $Q_g=n/\sigma^3$. Recall that the statistical theory predicts a constant value across small volume elements, which according Equation~(\ref{eq:Qacc}) is $Q_h/Q_g\approx 2$ for these particular experiments.  In contrast, the numerical halo models exhibit a noticeable variation of $Q_h$ as a function of distance from the point-mass. As $\epsilon$ decreases, the mean phase-space density plateaus at $Q_h/Q_g\lesssim 1$ at $r\lesssim r_\epsilon$, and $Q_h/Q_g\approx 3$ at $r\gtrsim r_\epsilon$, whereas the theoretical expectation matches the numerical curves at $r\simeq r_\epsilon$. This behaviour is consistent with the experiments run in \S\ref{sec:tests} with a much longer integration time, $t_f=100\,T(d)$.
 
In summary, Fig.~\ref{fig:con} indicates that for the range of point-masses explored in this paper the properties of the trapped haloes show converge when an empirical value of $\epsilon\lesssim 3 \times 10^{-3}$. The numerical experiments outlined in this work set $\epsilon=2\times 10^{-3}$.

 \begin{figure*}
\begin{center}
\includegraphics[width=170mm]{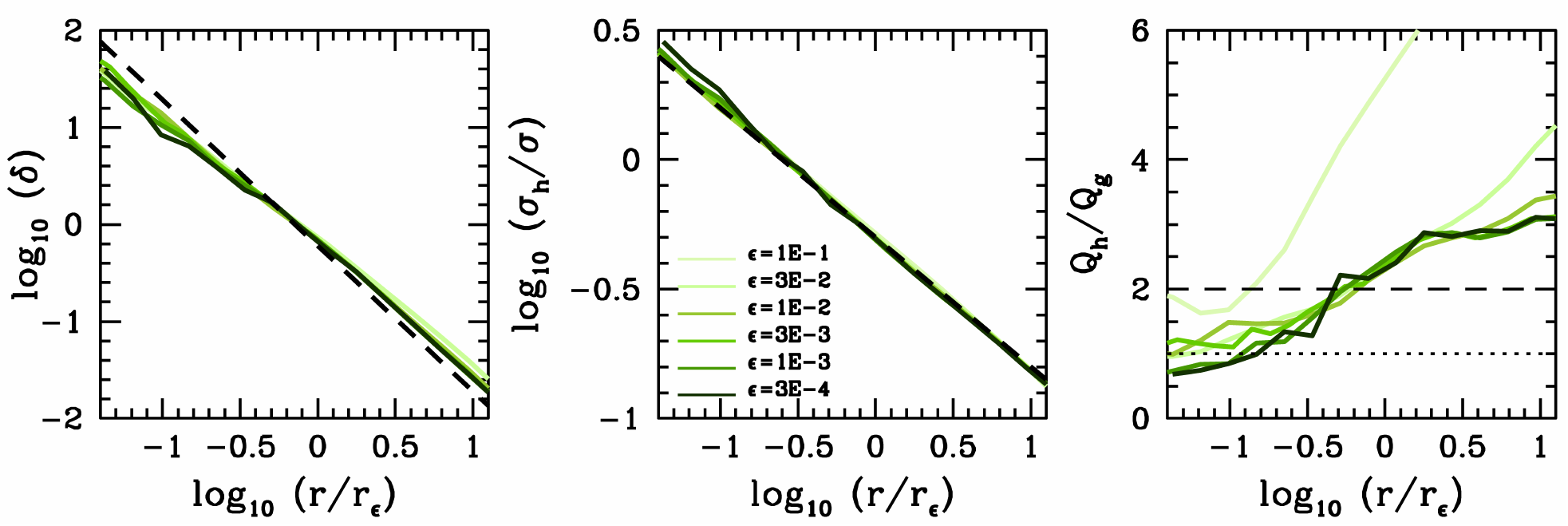}
\end{center}
\caption{Convergence tests. {\it Left panel}: Density enhancement of trapped halo particles as a function of distance from a point-mass $m_\star=3\times ^8 \msol$ for various choices of the parameter $\epsilon$. Distances are measured in units of the thermal critical radius, $r_\epsilon$, Equation~(\ref{eq:r_eps}). These models are run in a collisional mode ($F_\star\ne 0$). Black-dashed line plots the theoretical profile~(\ref{eq:delta}).  {\it Middle panel}: One-dimensional velocity dispersion profile associated with the models plotted in the left panel. Black-dashed line corresponds to the values predicted by Equation~(\ref{eq:vdisp}). {\it Right panel}: Mean phase-space density, $Q_h=\delta\,n/\sigma^3_h$, plotted in units of the local phase-space density of the field, $Q_g=n/\sigma^3$. The theoretical expectation for these particular models from Equation~(\ref{eq:Qacc}) is $Q_h/Q_g\approx 2$.}
\label{fig:con}
 \end{figure*}

\end{document}